\documentclass[lettersize,journal]{IEEEtran}
\usepackage{amsmath,amssymb,amsfonts}
\usepackage{algorithmic}
\usepackage{algorithm}
\usepackage{array}
\usepackage{stackengine}
\usepackage{textcomp}
\usepackage{stfloats}
\usepackage{url}
\usepackage{verbatim}
\usepackage{graphicx}
\usepackage{cite}
\usepackage{multirow}
\usepackage{subcaption}
\begin{document}

\title{A Laplacian Gaussian Mixture Model for Surface EMG Signals of Human Arm Activity}

\author{Durgesh Kusuru, \IEEEmembership{Student Member, IEEE}, Anish C. Turlapaty \IEEEmembership{Member, IEEE} and \\ Mainak Thakur  \IEEEmembership{Member, IEEE}
        % <-this % stops a space
\thanks{This research is funded by SERB, Govt. of India under Project Grant No. CRG/2019/003801.}% <-this % stops a space
\thanks{The authors are with the Bio-signal Analysis Group, Indian Institute of Information 
Technology Sri City, Chittoor, Andhra Pradesh, 517646, India. 
(e-mails: durgesh.k@iiits.in, anish.turlapaty@iiits.in and mainak.thakur@iiits.in}}

% % The paper headers
% \markboth{Journal of \LaTeX\ Class Files,~Vol.~14, No.~8, August~2021}%
% {Shell \MakeLowercase{\textit{et al.}}: A Sample Article Using IEEEtran.cls for IEEE Journals}

% \IEEEpubid{0000--0000/00\$00.00~\copyright~2021 IEEE}
% Remember, if you use this you must call \IEEEpubidadjcol in the second
% column for its text to clear the IEEEpubid mark.

\maketitle

\begin{abstract}
The probability density function (pdf) of surface Electromyography (sEMG) signals follows any one
of the standalone standard distributions: the Gaussian or the Laplacian. Further, the choice of the model is dependent on muscle contraction force (MCF) levels. Hence, a unified model is proposed which explains the statistical nature of sEMG signals at different MCF levels. In this paper, we propose the Laplacian Gaussian Mixture (LGM) model for the signals recorded from upper limbs. This model is able to explain the sEMG signals from different activities corresponding to different MCF levels. The model is tested on different bench-mark sEMG data sets and is validated using both the qualitative and quantitative perspectives. It is determined that for low and medium contraction force levels the proposed mixture model is more accurate than both the Laplacian and the Gaussian models. Whereas for high contraction force level, the LGM model behaves as a Gaussian model.  The mixing weights of the LGM model are analysed and it is observed that for low and medium MCF levels both the mixing weights of LGM model do contribute. Whereas for high contraction force levels the Laplacian weight becomes weaker. The proposed LGM model for sEMG signals from upper limbs explains sEMG signals at different MCF levels. The proposed model helps in improved understanding of statistical nature of sEMG signals and better feature representation in the classification problems.
\end{abstract}


\section{Introduction}
\subsection{Background}
{
Modeling of surface Electromyography (sEMG) signals has several applications such as 1) developing insights into sEMG signal generation from the constituent motor unit action potentials (MUAPs) that forms a basis for the sEMG signal synthesis \cite{wang2006simulation} and simulation studies  \cite{stashuk1993simulation}, 2) improving interpretation of the sEMG signals in clinical settings for example, in the diagnosis of neuromuscular disorders \cite{cuddon2002electrophysiology}, 3) analyzing inter-relations between the sEMG signals and the source muscle groups, for instance, in the sport sciences research \cite{vigotsky2018interpreting}, \cite{de1997use}, \cite{soderberg1984electromyography}, and 4) building visualization tools to support movement sciences \cite{hasanbelliu2004multi}, muscle physiology examinations and the sport science education.  
}
{
The sEMG signal models can be classified based on 1) bio-electrical, 2) statistical, and 3) machine learning principles. 
The earliest models were based on the physiological characteristics and the electrical activity in muscle fibers and motor units. For example, in \cite{merletti1999modeling} and \cite{rodriguez2012emg}, the sEMG signal is represented as a linear combination of MUAPs, where the action potential is modeled as a current tripole propagating from the neuromuscular junction to the fiber-tendon ending.}
In \cite{hayashibe2013voluntary}, a multi-scale physiological muscle model was used to estimate the muscle force from the sEMG signals corresponding to voluntary movements.

{
In the statistical approach, the sEMG signal is considered as a random signal and the typical characteristics modeled are the signal strength (samples), the temporal evolution of a signal, the autocorrelation of a single channel, and the spatial cross-correlations among multiple channels. The probabilistic models of the sEMG signal strength have evolved considerably during the last few decades as reviewed in the next section. In the temporal models, sEMG signals are usually represented by a linear autoregressive process \cite{paiss1987autoregressive}\cite{kiryu1994ar}. To estimate the MUAPs, the sEMG signals obtained from isometric contractions are modeled as an output of a LTI system with non-Gaussian white noise as an input \cite{shahid2005application}. In the variance based model, a sEMG signal is treated as a compound random process. For example, in a scale mixture model \cite{8627996}, the signal strength is modeled as a Gaussian process conditioned on the variance which is {modeled as an} inverse gamma variable.} 

{
The pattern classification of the sEMG signals plays a key role in applications such as the orthotic exoskeleton control\cite{khokhar2010surface}, the human movement analysis\cite{huang2008strategy}, and the neuromuscular disease diagnosis. For example, they can provide suitable inputs such as motor control parameters to drive a limb exoskeleton. In the machine learning methods, suitable features can be extracted based on the probability density function (pdf) of the sEMG signal \cite{chan2003continuous}. In the human movement analysis, sEMG signals can be used for discrimination among different actions, for example, hand gestures vs. grasping of objects \cite{furui2021emg}. In the neuormuscular disease diagnosis they can be used to study conditions such as myopathy which is related to the skeletal muscles causing them to become weaker and leading to muscle pain, weakness, fatigue and other symptoms\cite{giacomini2006electromyography}. Decoding information contained in the sEMG signals is critical and requires a reliable and precise solution. In human-machine interaction applications, deep learning methods play a crucial role and are used to achieve improved performance in tasks such as the movement classification, the joint angle prediction, and the force/torque estimation \cite{xiong2021deep,wei2019surface,hudgins1993new}}.
The focus of this paper is statistical modeling of the sEMG signal strength. 
\subsection{Existing Models for pdf of sEMG strength}\label{sec:LitRev}
{
Typical applications of a statistical signal model for sEMG signals are 1)  a better understanding of statistical nature of sEMG signals, 2) an improved feature representation in the classification problems, and 3) a qualitative analysis of signals. Depending on the muscle contraction level and the type of muscle, the existing models of sEMG signal strength are based on any of the standalone standard distributions such as the Gaussian or the Laplacian pdf. Following is a summary, based on studies since $1970s$, of the existing models of the sEMG signals acquired from different muscle groups of human upper limbs.}

{
In 1974, sEMG measurements were performed by Roesler
\cite{roesler1974statistical} and it was proposed that under constant force measurement conditions, the sEMG signals follow a Gaussian distribution. Miler-Brown et al. \cite{milner1975relation} observed that the distribution of the sEMG signals recorded from the first dorsal interosseus (FDI) muscle (back of a hand) at a lower force level has a sharper peak around zero than the Gaussian distribution and as the force level increases the sharpness near zero reduces. In \cite{parker1977signal}, the sEMG signals collected from biceps muscles were observed to follow a Gaussian distribution for the low and medium levels of MCF. Hunter et al. \cite{hunter1987estimation} analyzed the density of the sEMG signals from the biceps under constant MCF against a Gaussian density and reported that it has a narrow peak around zero. Later, Bilodeau et al. \cite{bilodeau1997normality} observed that for lower MCF levels, the sEMG signals from the biceps have a non-Gaussian nature with a peak near zero and at a higher MCF level their distribution was observed to tend toward a Gaussian model. Clancy and Hogan \cite{clancy1999probability} experimentally found that the density of sEMG signals at a constant MCF lies in between a Gaussian and a Laplacian pdf. In \cite{kaplanis2000bispectral}, it was noticed that the pdf of sEMG signal, 1)  has a sharper peak near zero and a longer tail than a usual Gaussian distribution at the low and high levels of MCF, and 2) follows a Gaussian model at a medium MCF level. In \cite{nazarpour2005negentropy}, at high MCF level, the distribution of the sEMG signals was found to be a Gaussian. Based on the recent studies, the sEMG signals at higher MCF levels from the flexor digitorum superficialis \cite{naik2011evaluation, naik2011kurtosis} and the biceps \cite{nazarpour2013note}, follow a Gaussian model.} 
%\textcolor{blue}
{
Based on this review, there is no unique statistical model that explains the activity at various contraction force levels. In many cases, it may not be possible to describe the data using the standard single density models. In such cases, often, modeling the data as a mixture of densities is an appropriate approach.
}
\textbf{Contributions}

%\textcolor{red}{Major contributions}

\begin{itemize}
    \item 
    %\textcolor{red}{A new mixture model for surface EMG signal amplitude:}\\
%\textcolor{blue}
{A unifying mixture model is proposed for the sEMG signals that explains the statistical nature of the signal for different levels of muscle contraction force. 
}
\item
%\textcolor{blue}
{The proposed model is tested on multiple benchmark sEMG datasets and the suitability of the model is compared against the existing models using both qualitative and quantitative methods. 
}
\item 
%\textcolor{blue}
{The weights of the mixture components are analyzed for different activities and intensities
and a possible inter-relation is illustrated. 
}
\end{itemize}

\section{Statistical Model and Problem Description}
\subsection{Laplacian Gaussian Mixture Model}
%\textcolor{red}{Precise meaning of modeling - eg. here we are modeling the amplitude of surface EMG signals} In this paper, a statistical model is proposed for the sEMG signal samples.. 
%\textcolor{blue}
{In \cite{kusuru2021laplacian}, a Laplacian Gaussian Mixture (LGM) model was introduced and verified on a single sEMG dataset. In this work, the LGM model is further analyzed and its suitability is evaluated for various benchmark datasets corresponding to distinct upper limb activities at different MCF levels. A description of the proposed model follows.}

{
Let the strength of the discrete time sEMG signal be represented by a random variable $Y$. The LGM model is written as
\begin{equation} \label{model:LGM1}
f_{Y}(y;\Theta )= \lambda{_1}f_1(y;\theta{_1}) + \lambda{_2} f_2(y;\theta{_2})
\end{equation}
$y$ denotes a realization of $Y$ and $\Theta = [\lambda_1, \lambda_2, \theta_1, \theta_2]$ is the set of unknown parameters. $\lambda_1$ and $\lambda_2$ are the mixing weights that add to unity. $\theta_1$ and $\theta_2$ are parameters of component densities. $f_1(y;\theta_1)$ is a Laplacian density defined as 
\begin{equation}\label{Comp:L}
f_{1}(y; \theta{_1}) = \frac{1}{2\sigma{_1}} \exp \bigg(-\frac{\left | y-\mu{_1} \right |}{\sigma{_1}}\bigg)   ~   -\infty  < y  < \infty
\end{equation}
and $f_2(y;\theta_2)$ a Gaussian density given by
\begin{equation}\label{Comp:G}
f_{2}(y; \theta{_2})= \frac{1}{\sqrt{2\pi\sigma_2^2}} \exp \bigg(- \frac{(y-\mu_2)^2}{2\sigma_2^2}\bigg)   ~   -\infty  < y < \infty
\end{equation}
 note that $\theta_1 = [\mu_1,\sigma_1 ]$ ~ and ~$\theta{_2} = [\mu_2, \sigma^2_2]$  {are {parameters} of the respective densities.}}
As illustrated in (\ref{model:LGM1}), the mixing weights $\lambda_1$ and $\lambda_2$ are the hidden parameters. The unknown parameters of the LGM model are estimated from the sEMG data using the expectation-maximization (EM) Algorithm\cite{bishop2006pattern}. Note that the EM algorithm is commonly used for estimation of parameters of the Gaussian mixture model based on which a similar EM methodology is derived for the proposed model.

\subsection{{ Parameter Estimation Problem}}
{ Consider an array $\textbf{y}= \{y_n\}_{n=0}^{N-1}$ where $y_n$ represents a discrete sample of a sEMG signal. Based on the latent variable used in Gaussian mixture models \cite{bishop2006pattern}, a discrete random vector \textbf{w} is defined as 
\begin{equation}
 \textbf{w} = \{w_n\}_{n=0}^{N-1}  
\end{equation}
here  $w_n = [w_{n,1}, w_{n,2}] $  and has two distinct states with corresponding likelihoods (mixing weights)
%\begin{eqnarray}
%  \big \{\{w_{n,1} = 1, w_{n,2} = 0\}, \{w_{n,1} = 0, w_{n,2} = 1\} \big\}   
%\end{eqnarray}
%The relation with the mixing coefficients is emphasized by the likelihoods
\begin{eqnarray}
p(w_{n,1} = 1, w_{n,2} = 0) &=& \lambda_1  \\
p(w_{n,1} = 0, w_{n,2} = 1) &=& \lambda_2  \nonumber
\end{eqnarray}
and the marginal likelihood of these hidden states is given by
\begin{equation}
 p(w_n) = \lambda^{w_{n,1}}_1 \lambda^{w_{n,2}}_2
\end{equation}
%The variables $w_n$ are assumed to be i.i.d., hence
The conditional pdf of $y_n$ given $w_n$ and $\Theta$ is 
\begin{equation}  \label{cond_pdf_zn}
 f(y_n|w_n;\Theta)= \prod_{j=1}^{2}(f_j(y_n;\theta_j))^{w_{n,j}}
\end{equation}
%conditionally
Here, $y_n$ are i.i.d. The joint density of the data, the hidden states and the unknown parameters is
\begin{equation}
f(\textbf{y},\textbf{w};\Theta)=  \prod_{n=0}^{N-1} \prod_{j=1}^{2} (\lambda_{j}f_{j}(y_{n};\theta_{j}))^{w_{n,j}}
\end{equation}
The estimation problem can be stated as follows: given the data \textbf{y} which follows the LGM model (\ref{model:LGM1}), the objective is to estimate the parameters
$\Theta$ and the related statistics in the model (\ref{model:LGM1}). {The next section describes the parameter estimation for the LGM model using the EM algorithm.}
}
\subsection{{ EM-Algorithm}}
{
The complete data log-likelihood is
\begin{equation}
    L(\textbf{y},\textbf{w};\Theta)= \sum_{n=0}^{N-1} \sum_{j=1}^{2}w_{n,j} \ln(\lambda_{j}f_{j}(y_n;\theta_j))
\end{equation} 
\subsubsection{E-step}
 Given the data $\textbf{y}$ and the recent estimate of $\Theta$ represented by $\Theta^{(i)}$, $\Lambda(\textbf{y},\Theta ,\Theta^{(i)})$ is the expectation of the full data log-likelihood evaluated with respect to the conditional likelihood of hidden variables.
 \begin{equation}
     \Lambda(\textbf{y},\Theta ,\Theta^{(i)})=E_{\textbf{w}|\textbf{y},\Theta^{(i)}} \big\{L(\textbf{y}, \textbf{w};\Theta)\big\}
 \end{equation}
 The posterior probability of $w_n$ is evaluated using Bayes theorem as
 
\begin{equation} \label{postw_nj}
    P(w_{n,j}=1|y_n;\Theta^{(i)})=\frac{f(y_{n}|w_{n,j}=1;\theta_j^{(i)})P(w_{n,j}=1)}
    {\sum_{l=1}^{2}f(y_{n}|w_{n,l}=1;\theta_j^{(i)})P(w_{n,l}=1)}
\end{equation}
note that the Bayesian estimate of $w_n$ is
\begin{equation} \label{E_w_nj}
E(w_{n}|y_n,\Theta^{(i)})=P(w_{n,j}=1|y_n,\theta_j^{(i)})
 \end{equation}
 based on (\ref{cond_pdf_zn}), for $w_{n,j} = 1$ the conditional pdf $f(y_n|w_n;\Theta)$ reduces to a component density. 
 Then the estimate (\ref{E_w_nj}), denoted by $\gamma^{(i)}_{n,j}$, can be written as 
\begin{equation}
    \gamma^{(i)}_{n,j} =\frac{\lambda_j f_{j}(y_{n};\theta_j^{(i)})}{\sum_{i=1}^2\lambda_i f_{i}(y_{n};\theta^{(i)}_i)}
\end{equation}
Thus, the expectation on the complete data log likelihood becomes
\begin{equation} \label{exp:DLH}
   \Lambda(\textbf{y},\Theta,\gamma^{(i)})=\ \sum_{i=1}^{n} \sum_{j=1}^{2}\gamma^{(i)}_{n,j} \ln(\lambda_j f_j(y_n;\theta_j))
\end{equation}}
{where 
\begin{equation}
    \gamma^{(i)}=\{\gamma^{(i)}_{0,1}, \gamma^{(i)}_{2,1},...,\gamma^{(i)}_{N-1,1}, \gamma^{(i)}_{0,2}, \gamma^{(i)}_{1,2},..., \gamma^{(i)}_{N-1,2}\}
\end{equation}}

%To determine the parameter updates $\Theta^{(i+1)}$, the corresponding
\subsubsection{M-step}
%\textcolor{blue}
{
Substituting both the Laplacian pdf (\ref{Comp:G}) and the Gaussian  pdf (\ref{Comp:L}) in (\ref{exp:DLH}) leads to
\begin{eqnarray} \label{DataLike}
 \Lambda(\textbf{y},\Theta,\gamma^{(i)}) &=&  \sum_{n=0}^{N-1} \gamma^{(i)}_{n,j}  \bigg\{ \ln \lambda_1 - \ln \sigma_1 - \frac{|y_n - \mu_1|}{ \sigma_1} \nonumber \\ &&
 \ln \lambda_2 - \frac{1}{2} \ln \sigma^2_2 - \frac{(y_n - \mu_2)^2}{2 \sigma^2_2}     \bigg\}
\end{eqnarray}
Based on the optimization problem given below, the parameters are estimated iteratively.
\begin{equation}
    \Theta^{(i+1)} =  \max_{\Theta}~ \Lambda(\textbf{y},\Theta,\gamma^{(i)})
\end{equation}
%By substituting the data likelihood in  for the $\Lambda(\vy,\Theta,\gamma{(i)})$. 
By equating the partial derivatives of $\Lambda(\textbf{y},\Theta,\gamma^{(i)})$ in (\ref{DataLike}) to zero and solving the corresponding equations, the estimates of the parameters
are obtained as follows. 
\begin{eqnarray} \label{Par_Updates}
    \lambda^{(i+1)}_1&=&\frac{N_1}{N}  \nonumber \\
    \lambda^{(i+1)}_2&=&\frac{N_2}{N} \nonumber \\
     \mu^{(i+1)}_1&=&  Median \bigg[ \bigg\{\frac{\gamma^{(i)}_{n,1}}{N_1},  y_n \bigg\}_{n=0}^{N-1}\bigg] \nonumber \\
   (\sigma_{1})^{(i+1)} & = &  \frac{1}{N_1}  {\sum_{n=0}^{N-1}\gamma^{(i)}_{n,1} \left | (y_n-\mu^{(i)}_1)\right |}   \\
    \mu^{(i+1)}_2&=& \frac{1}{N_2} \sum_{n=0}^{N-1}\gamma^{(i)}_{n,2} y_n  \nonumber  \\
    (\sigma^{2}_{2})^{(i+1)}&=& \frac{1}{N_2} {\sum_{n=0}^{N-1}\gamma^{(i)}_{n,2}(y_n-\mu^{(i)}_2)^2}  \nonumber
\end{eqnarray}
where $N_1 = \sum_{n=0}^{N-1}\gamma^{(i)}_{n,1}$ and $N_1 + N_2 = N$.
{The E \& M steps are iterated until the squared difference between two successive estimates $\Theta^{(i)}$ and $\Theta^{(i+1)}$ converges.}
}
\subsection{Evaluation Methods}
%\textcolor{blue}
{
{The parameter estimates from the EM algorithm (\ref{Par_Updates}) are used to generate a fit of the LGM pdf for the sEMG samples as follows}
\begin{equation}
    {f}(y;\hat{\Theta} )= {{\hat{\lambda}}_1}f_1(y; {\hat{\mu}}_1,{\hat{\sigma}}_1) + {{\hat{\lambda}}_2}f_2(y; {\hat{\mu}}_2,{\hat{\sigma}}^2_2)
\end{equation}
here, ${{\hat{\lambda}}_1}, {\hat{\mu}}_1, {\hat{\sigma}}_1, {{\hat{\lambda}}_2}, {\hat{\mu}}_2, {\hat{\sigma}}^2_2 $ are the estimates from 
(\ref{Par_Updates}) at convergence. 
The empirical pdf (mpdf) is constructed from the histogram of the signal samples. {The evaluation criteria for the appropriateness of the model are mentioned below}
}

\textbf{{Visual inspection}}:
{
The model based pdf i.e. the approximate pdf fitted from a model and the mpdf are compared visually for understanding the degree of agreement \cite{spanos2019probability}.}  

\textbf{Kullback–Leibler divergence}: 
{
Kullback–Leibler divergence(KLD)\cite{kullback1997information} is a statistical metric that measures the difference between two pdfs. Let $p_{1}$ and $p_{2}$ be two probability densities then the KLD between them is 
\begin{equation}
D_{KL}(p_1 || p_2) =\sum_{x}p_1(x) \ln \bigg(\frac{p_1(x)}{p_2(x)} \bigg)
\end{equation}\\
in this paper, {$p_1$ is the empirical distribution and $p_2$ is a model based approximate pdf}. If {these} two distributions match then the $D_{KL}(p_1 || p_2)$  equals $0$. The lower the $D_{KL}(p_1 || p_2)$, the closer the approximation is to the mpdf.}
\begin{table*}[t]
\centering
\caption{Basic characteristics of four benchmark sEMG datasets}
\begin{tabular}{c c c c c}\\ 
                           \hline \hline & Ninapro DB2                                                                                               & Ninapro DB4                                                                                                 & Rami-khushaba DB6 & Intense Dataset \\
\hline No. of Subjects              & 40                                                                                                        & 10                                                                                                          & 11                & 15              \\
\hline Total no. of activities      & \begin{tabular}[c]{@{}c@{}}Exercise-1- 17\\ Exercise-2- 23\\ Exercise-3- 09\\       Total 49\end{tabular} & \begin{tabular}[c]{@{}c@{}}Exercise-1-   12\\ Exercise-2- 17\\ Exercise-3-  23\\      Total 52\end{tabular} & 40                & 1               \\
\hline No. of activities considered & 23                                                                                                        & 17                                                                                                          & 40                & 1               \\
\hline No. of repetitions           & 6                                                                                                         & 6                                                                                                           & 6                 & 1               \\
\hline No. of channels              & 12                                                                                                        & 12                                                                                                          & 7                 & 8               \\
\hline Type of electrode            & Delsys                                                                                                    & Cometa Mini Wave                                                                                            & Delsys            & Myo-armband     \\
\hline Sampling rate                & 2000 samples/sec                                                                                          & 2000 samples/sec                                                                                  & 4000samples/sec  

              & 200 samples/sec\\
              \hline \hline
\end{tabular} 
\label{Datasets summary}
\end{table*}

\begin{figure*}[h]
\captionsetup[subfigure]{justification=centering}
    \centering
      \begin{subfigure}{0.31\textwidth}
        \includegraphics[width=\textwidth]{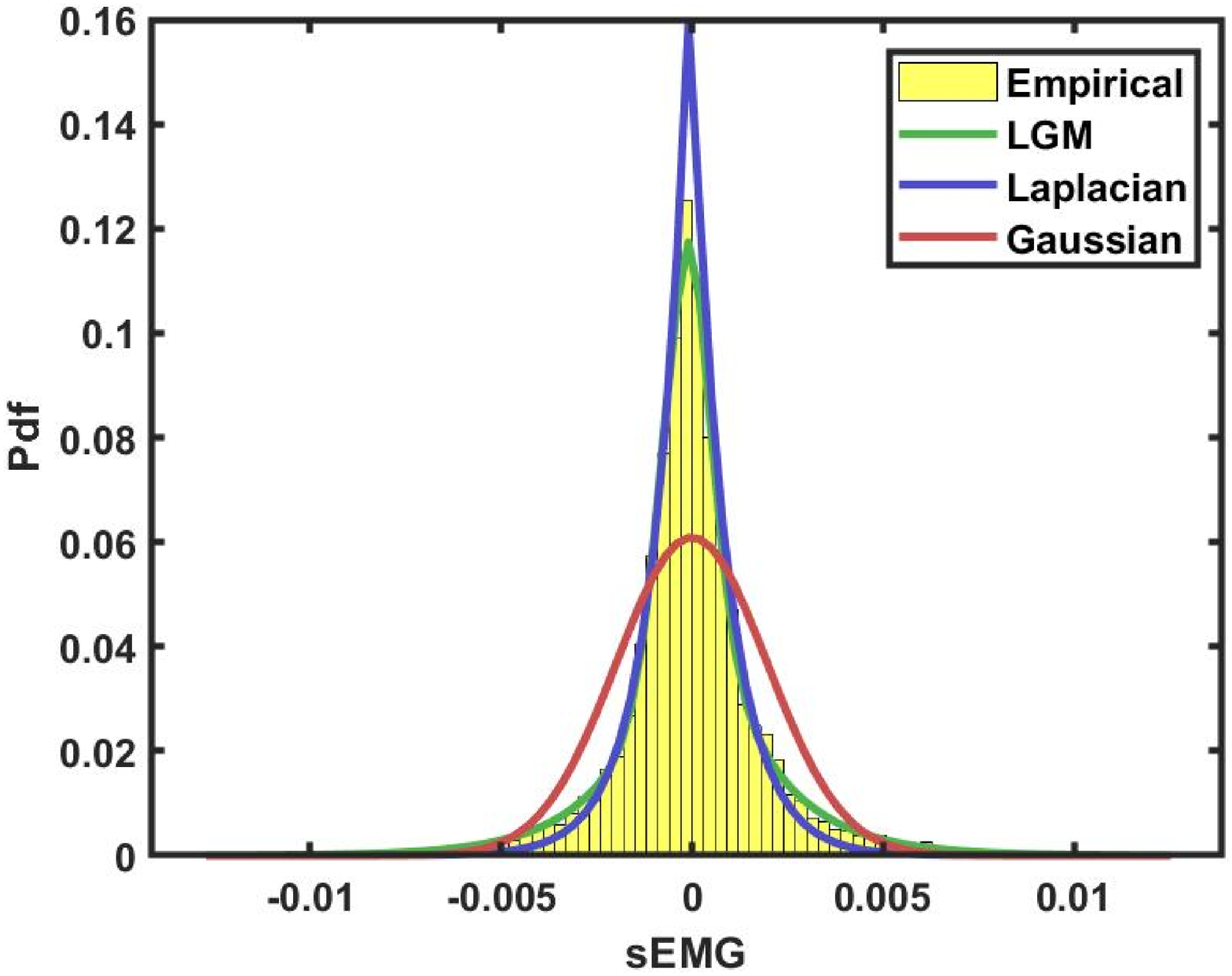}
          \caption{Gestures}
          \label{fig:NiceImage1}
      \end{subfigure}
      \begin{subfigure}{0.31\textwidth}
        \includegraphics[width=\textwidth]{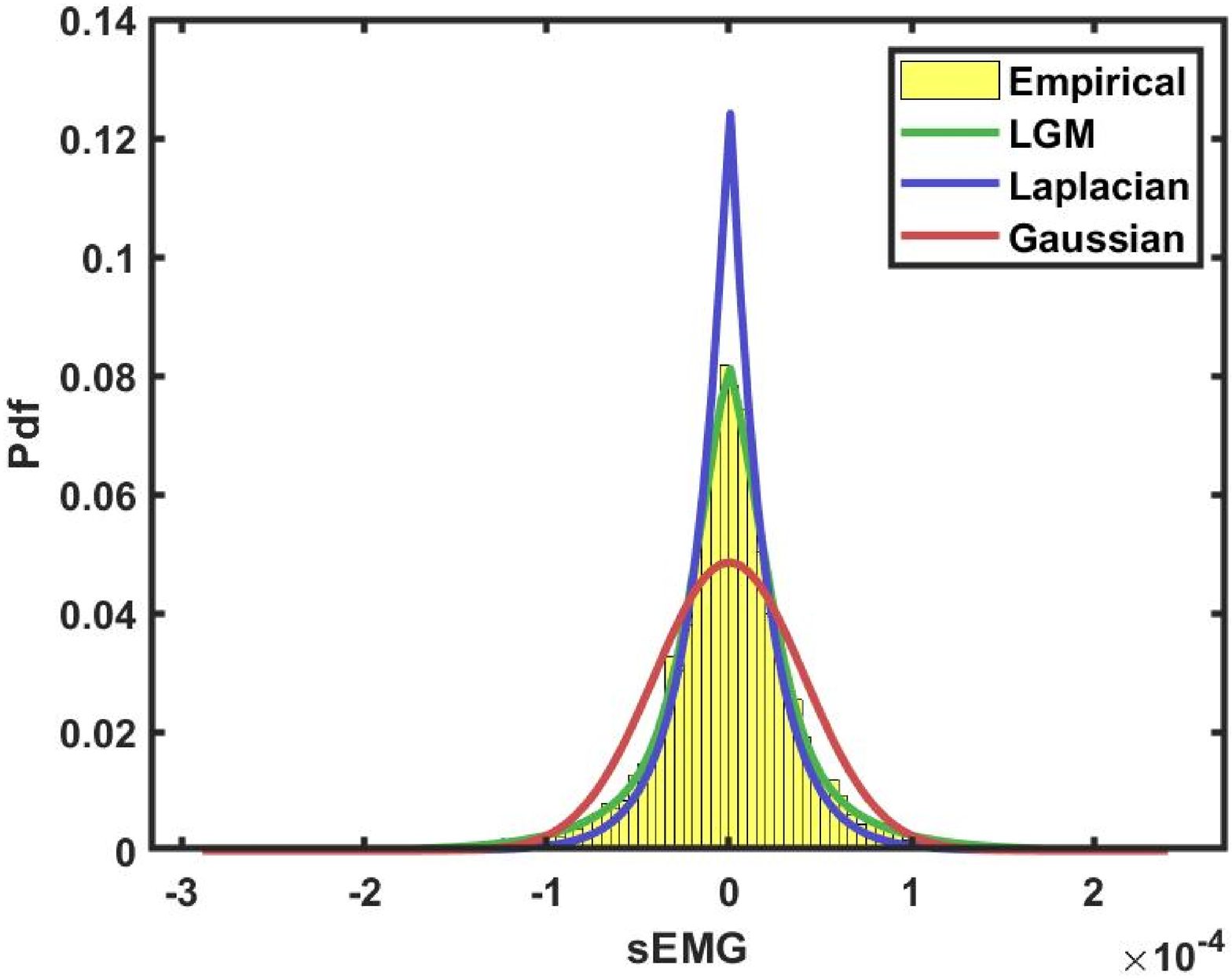}
          \caption{Grasping}
          \label{fig:NiceImage2}
      \end{subfigure}
     
      \begin{subfigure}{0.31\textwidth}
        \includegraphics[width=\textwidth]{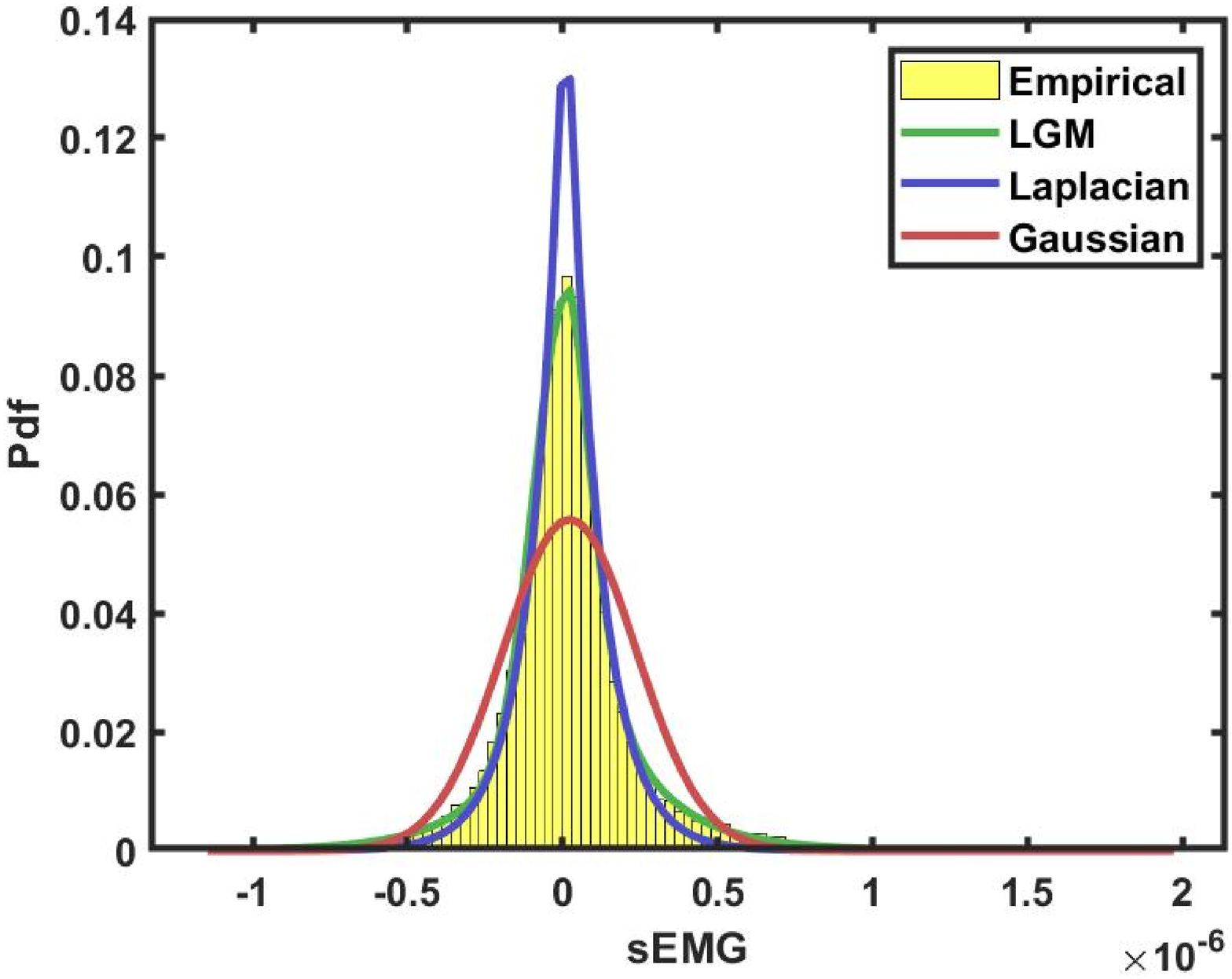}
          \caption{Arm activity}
          \label{fig:NiceImage1}
      \end{subfigure}
      \begin{subfigure}{0.31\textwidth}
        \includegraphics[width=\textwidth]{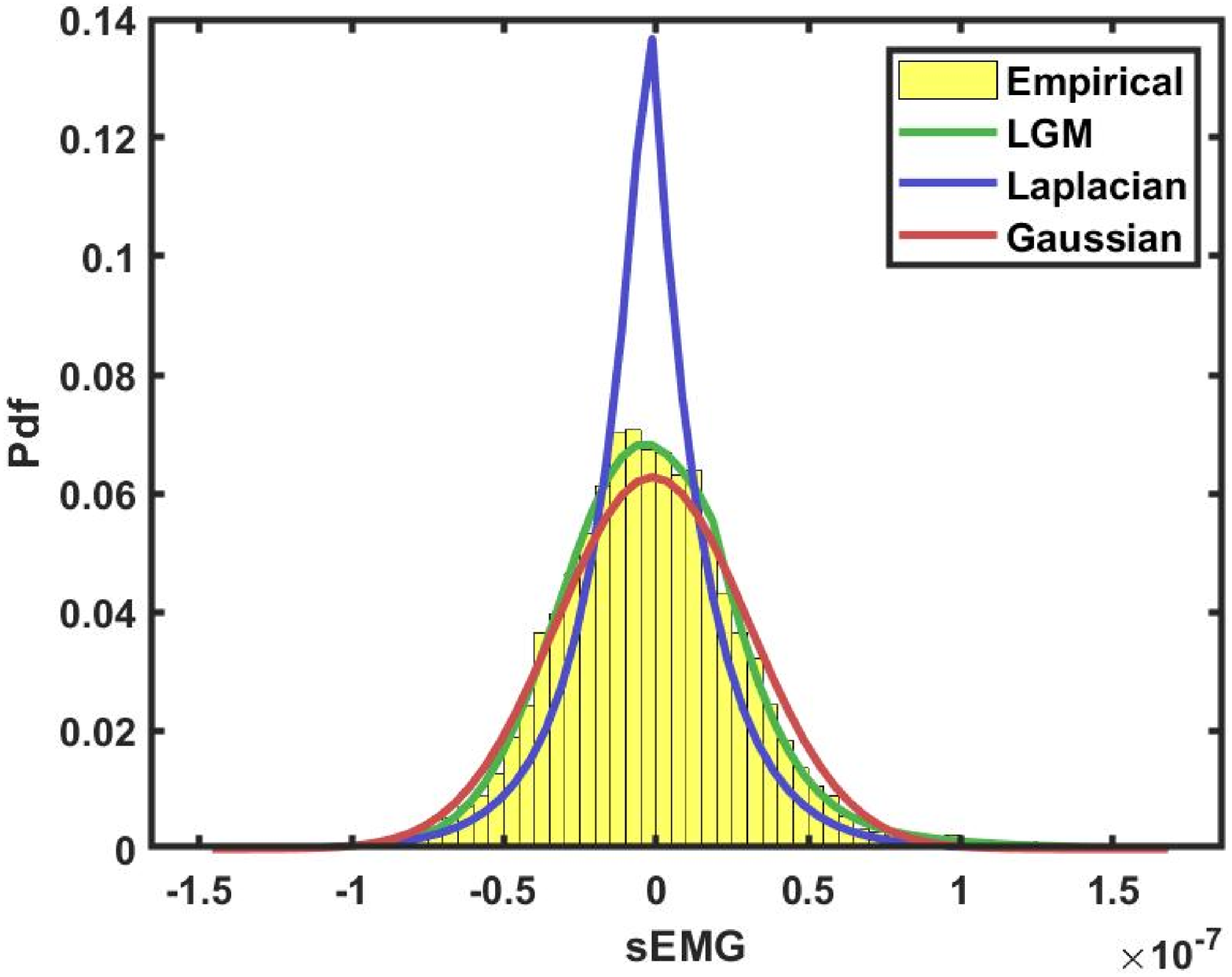}
          \caption{Intense activity}
          \label{fig:NiceImage2}
      \end{subfigure}
      
     \caption {Visual comparisons between mpdfs and estimated pdfs from models: LGM(green), Laplacian(blue) and Gaussian(red) for gestures, grasping, arm and intense activities for the subjects - 10, 3, 1 and  10 with corresponding activities - 7, 18, 3 and 1}
      \label{fig:VisualComp}
\end{figure*}

\begin{figure*}[h]
\captionsetup[subfigure]{justification=centering}
    \centering
      \begin{subfigure}{0.31\textwidth}
        \includegraphics[width=\textwidth]{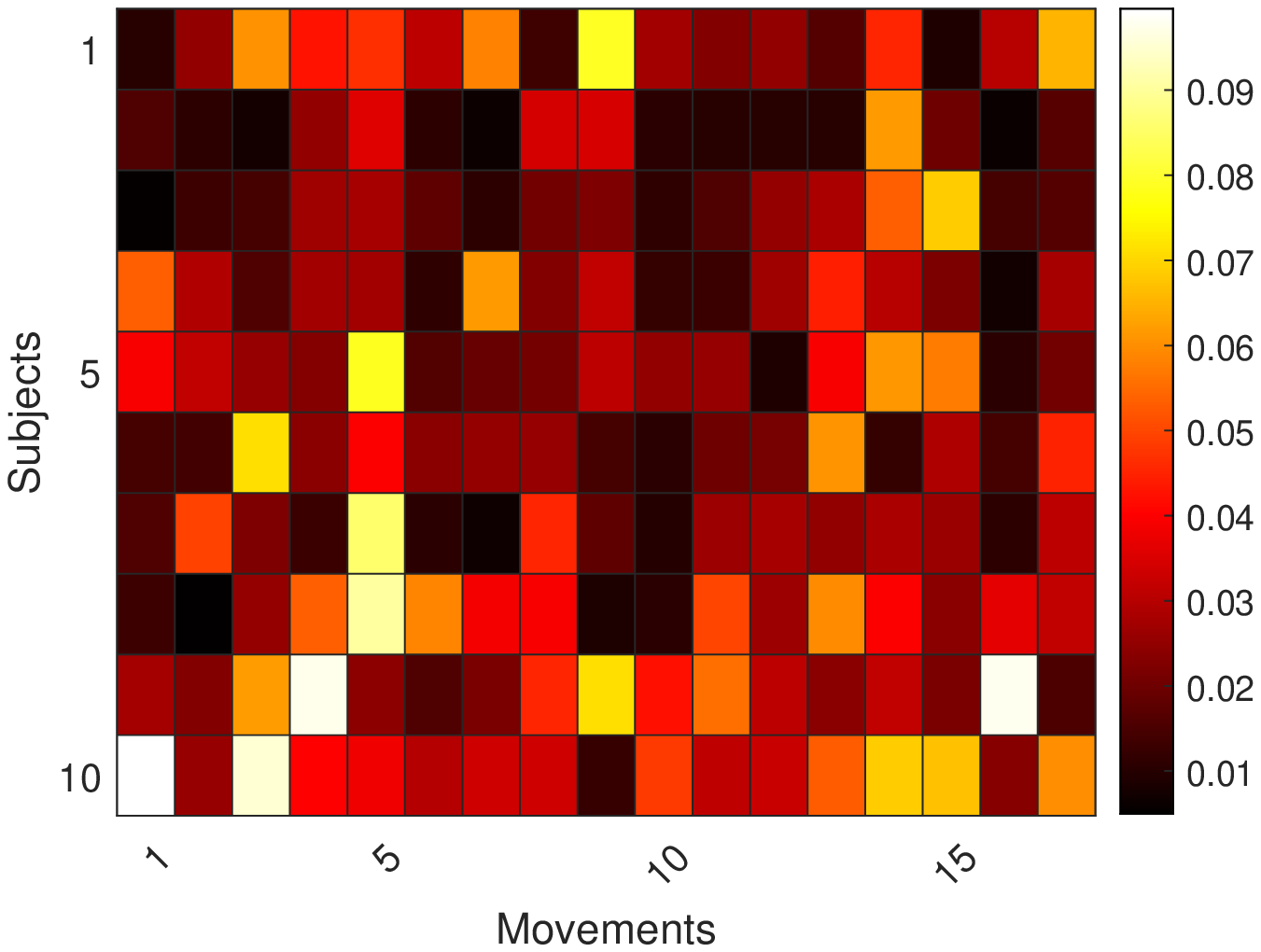}
          \caption{}
          \label{fig:NiceImage1}
      \end{subfigure}
      \begin{subfigure}{0.31\textwidth}
        \includegraphics[width=\textwidth]{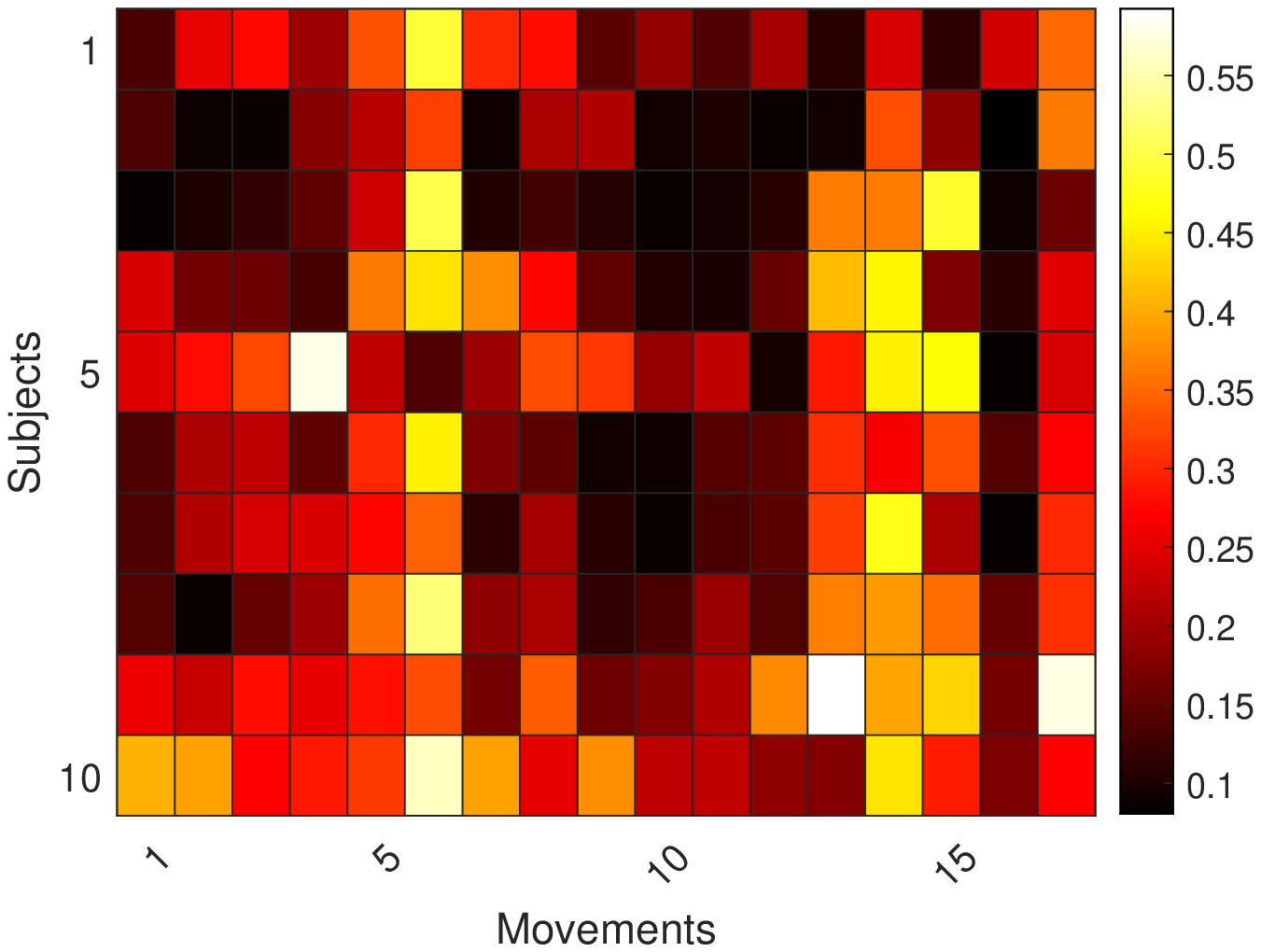}
          \caption{}
          \label{fig:NiceImage2}
      \end{subfigure}
      \begin{subfigure}{0.31\textwidth}
        \includegraphics[width=\textwidth]{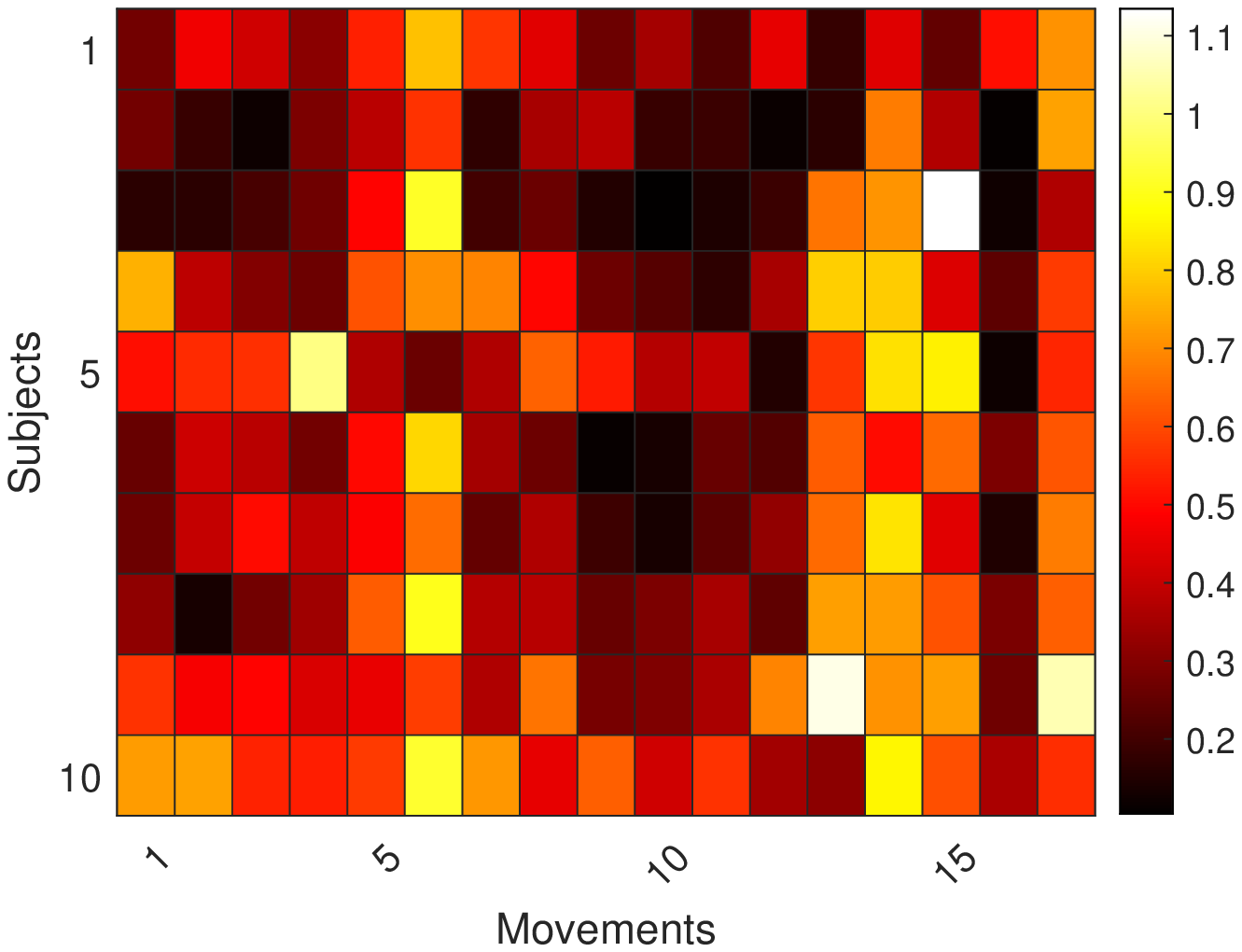}
          \caption{}
          \label{fig:NiceImage3}
      \end{subfigure}
      \begin{subfigure}{0.31\textwidth}
        \includegraphics[width=\textwidth]{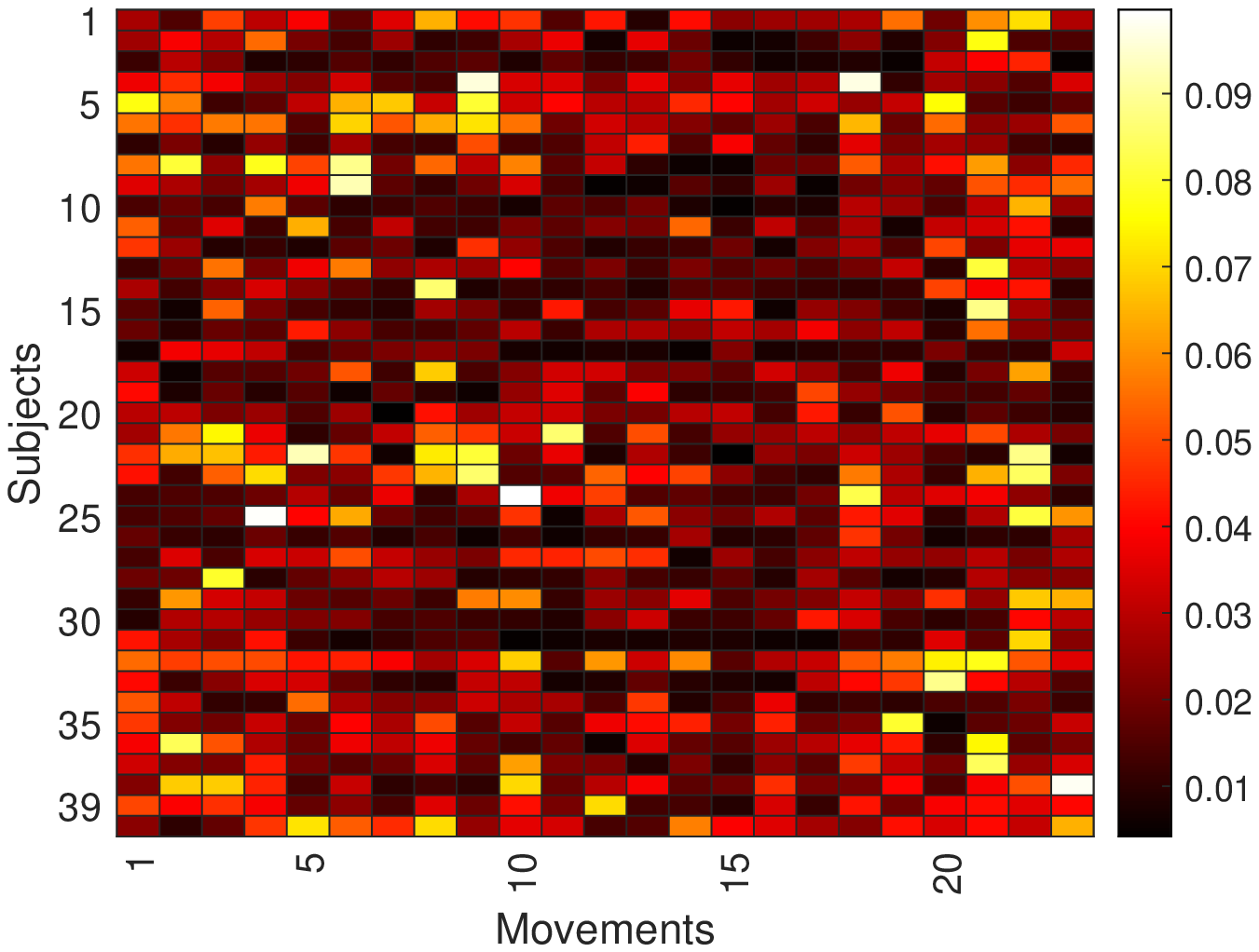}
          \caption{}
          \label{fig:NiceImage1}
      \end{subfigure}
      \begin{subfigure}{0.31\textwidth}
        \includegraphics[width=\textwidth]{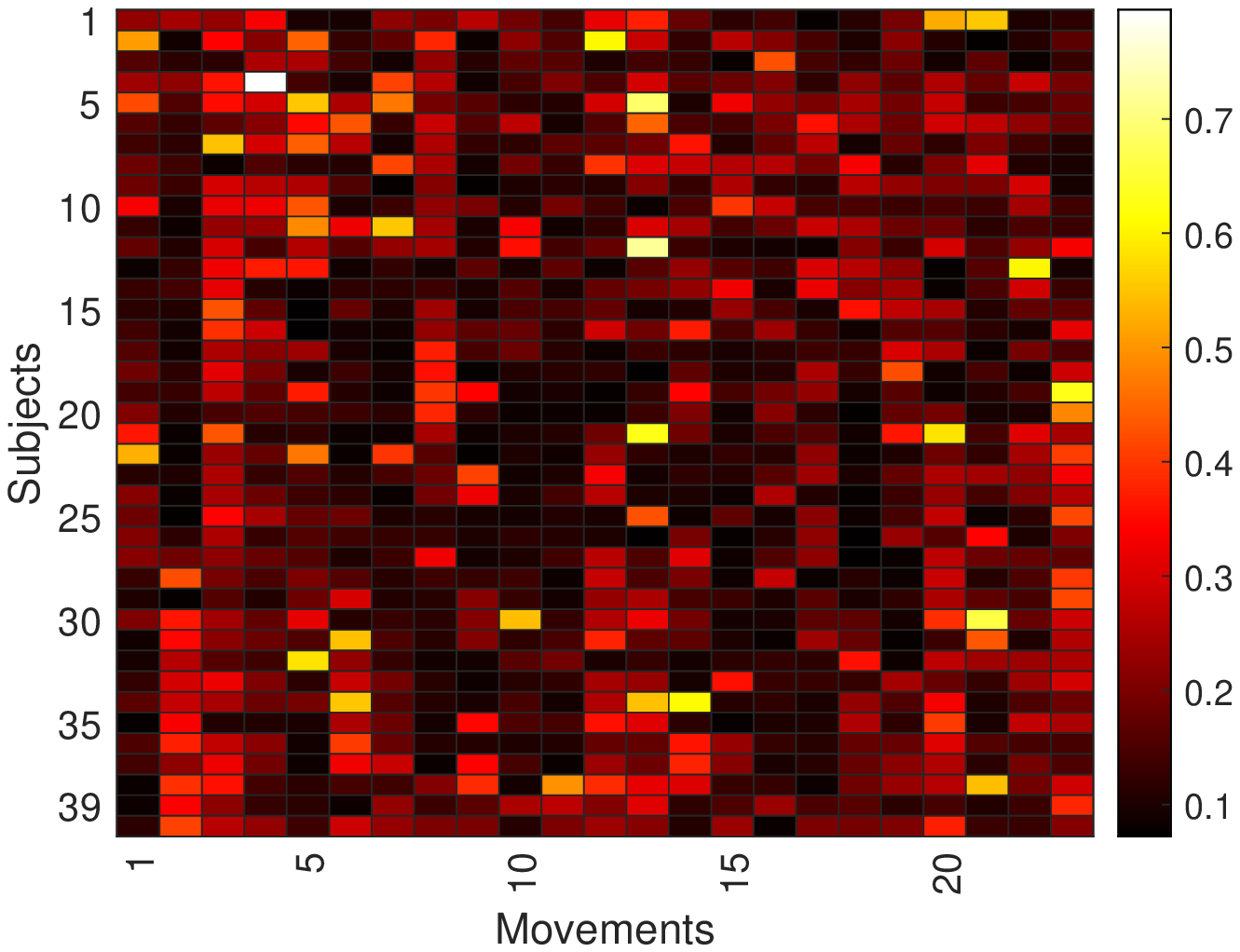}
          \caption{}
          \label{fig:NiceImage2}
      \end{subfigure}
      \begin{subfigure}{0.31\textwidth}
        \includegraphics[width=\textwidth]{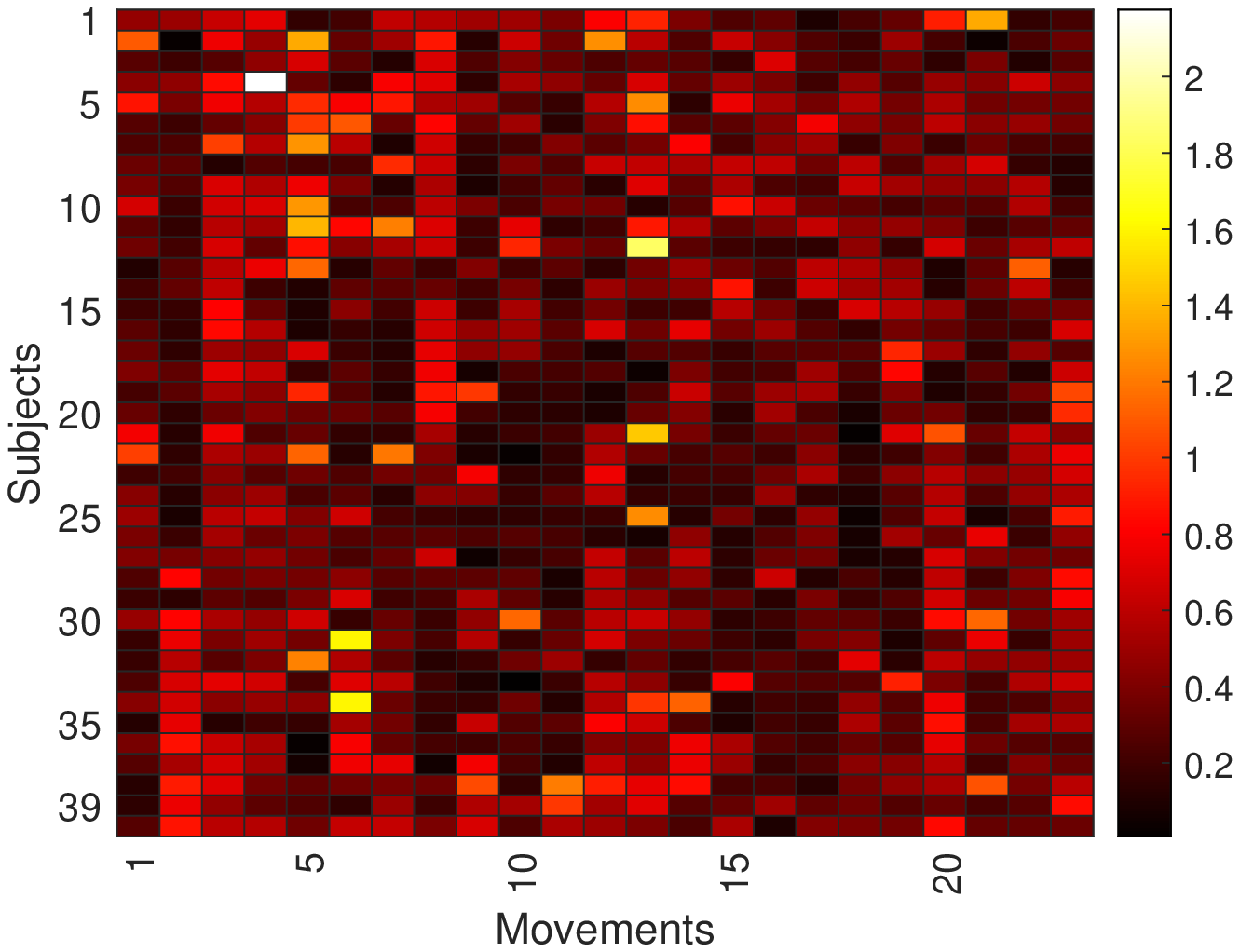}
          \caption{}
          \label{fig:NiceImage3}
      \end{subfigure}
      \begin{subfigure}{0.31\textwidth}
        \includegraphics[width=\textwidth]{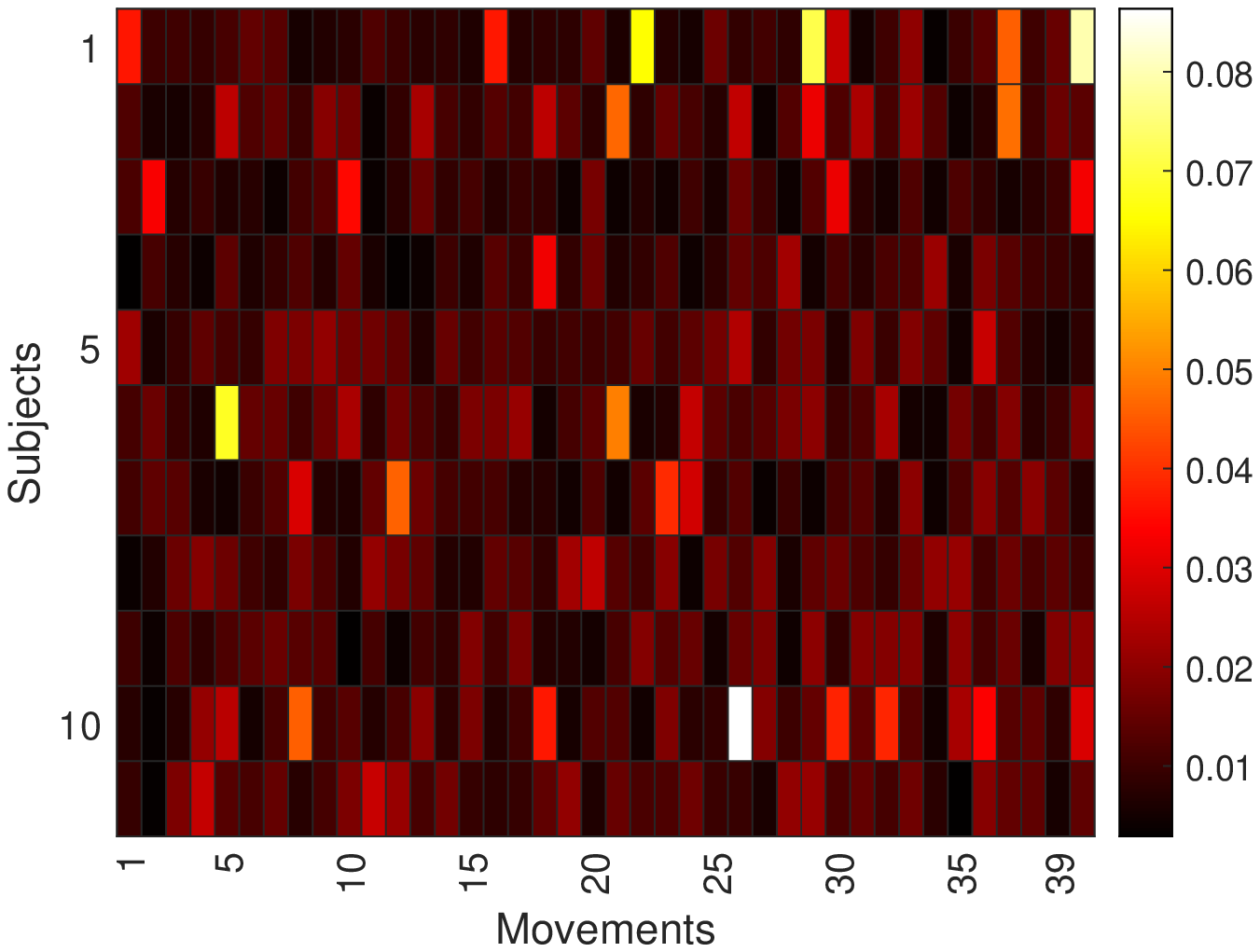}
          \caption{}
          \label{fig:NiceImage1}
      \end{subfigure}
      \begin{subfigure}{0.31\textwidth}
        \includegraphics[width=\textwidth]{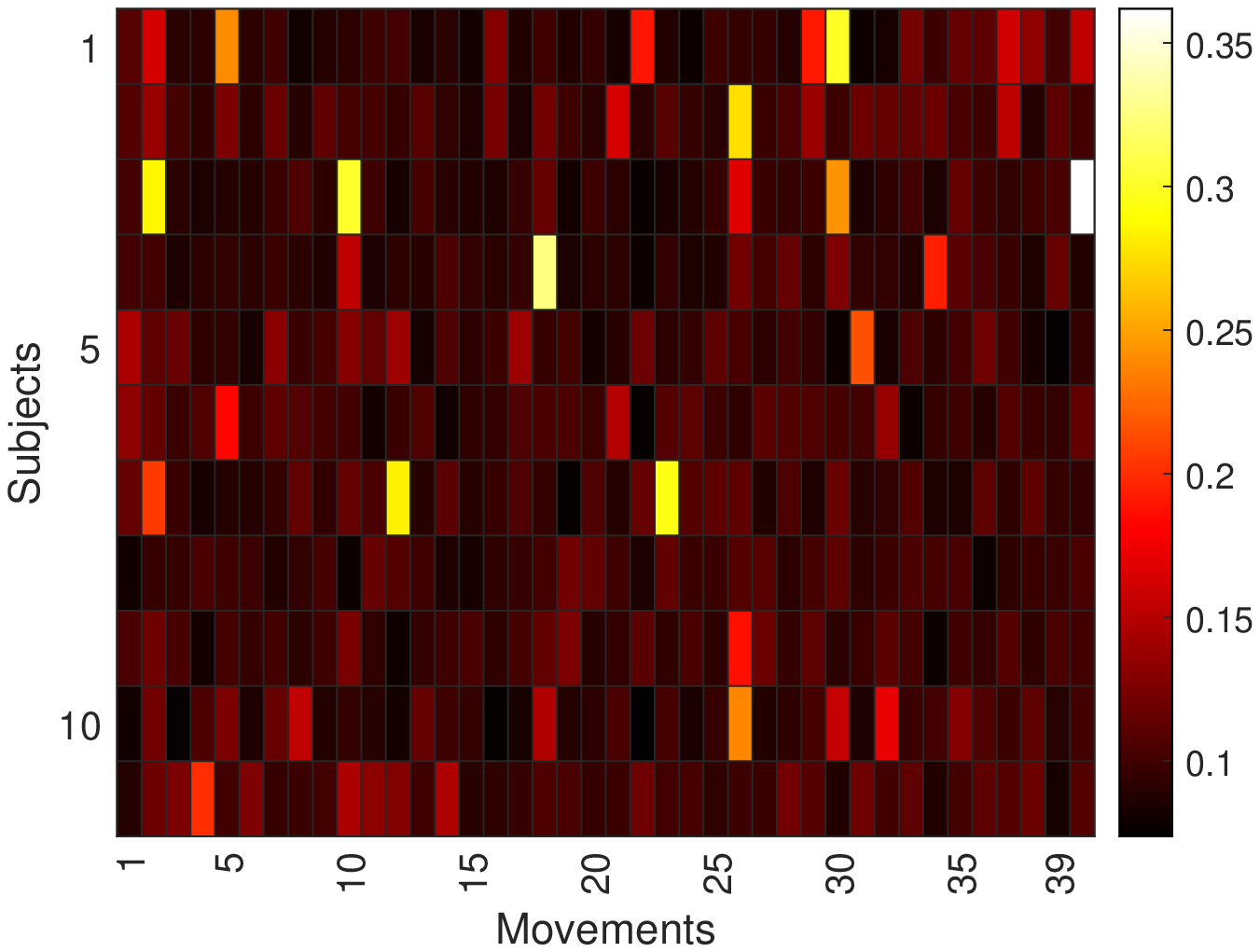}
          \caption{}
          \label{fig:NiceImage2}
      \end{subfigure}
      \begin{subfigure}{0.31\textwidth}
        \includegraphics[width=\textwidth]{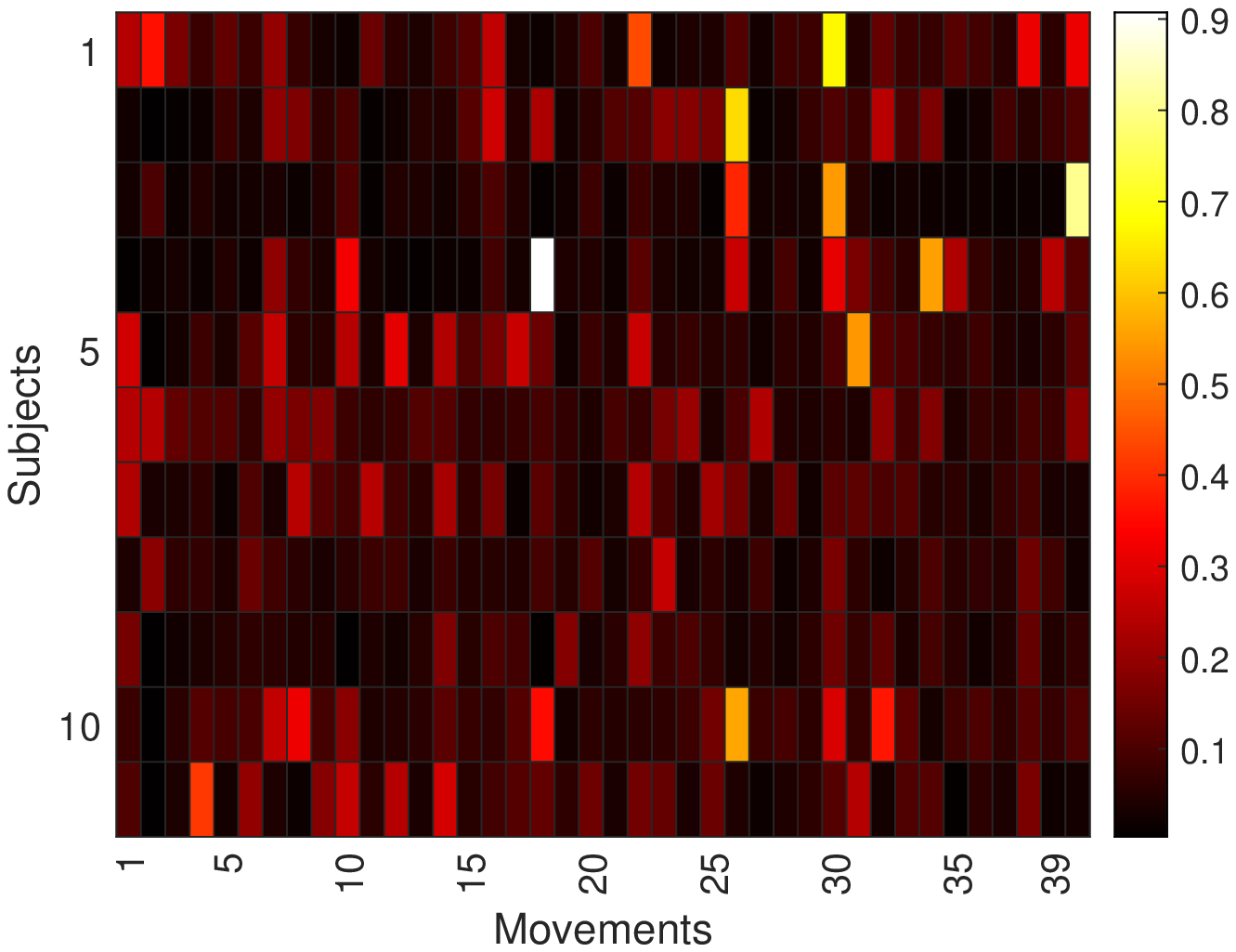}
          \caption{}
          \label{fig:NiceImage3}
      \end{subfigure}
      \caption{Heatmaps of KLD for the 3 models: (a) LGM, (b) Laplacian and (c) Gaussian corresponding to Ninapro-DB4, (d) LGM, (e) Laplacian and (f) Gaussian from Ninapro-DB2 and (g) LGM, (h) Laplacian and (i) Gaussian from Rami-khushaba-DB6}
      \label{heatmap}
\end{figure*}
\begin{figure}[!t]
        \centering
      \includegraphics[width=0.91\columnwidth]{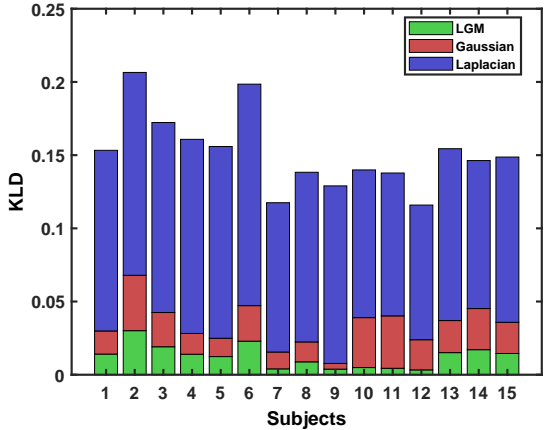}
      \caption{KLD values of LGM, Laplacian and Gaussian models for intense activity data}
     \label{intense}
\end{figure}

\begin{figure*}[htb]
\captionsetup[subfigure]{justification=centering}
    \centering
      \begin{subfigure}{0.45\textwidth}
        \includegraphics[width=\textwidth]{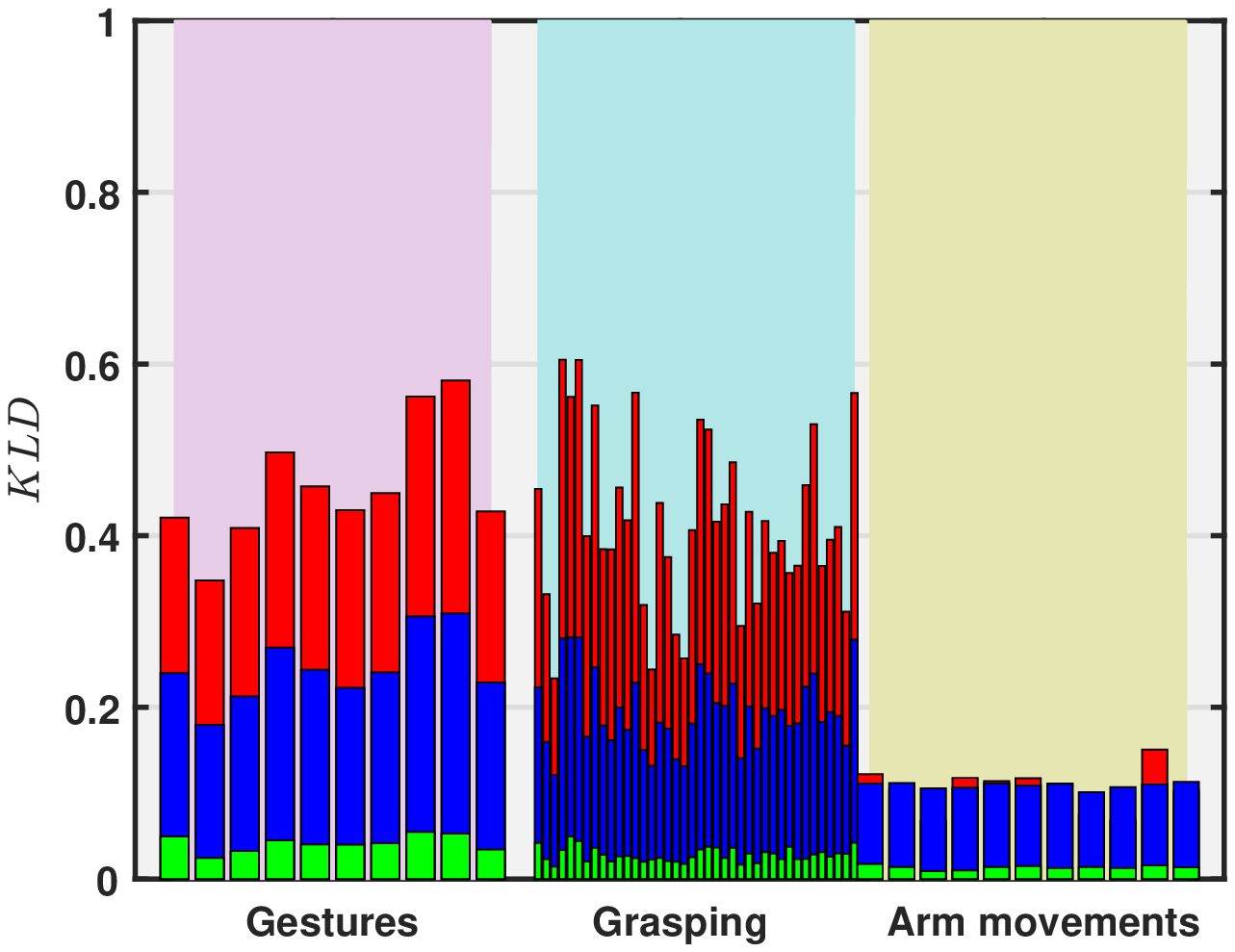}
          \caption{}
          \label{fig:NiceImage1}
      \end{subfigure}
      \begin{subfigure}{0.45\textwidth}
        \includegraphics[width=\textwidth]{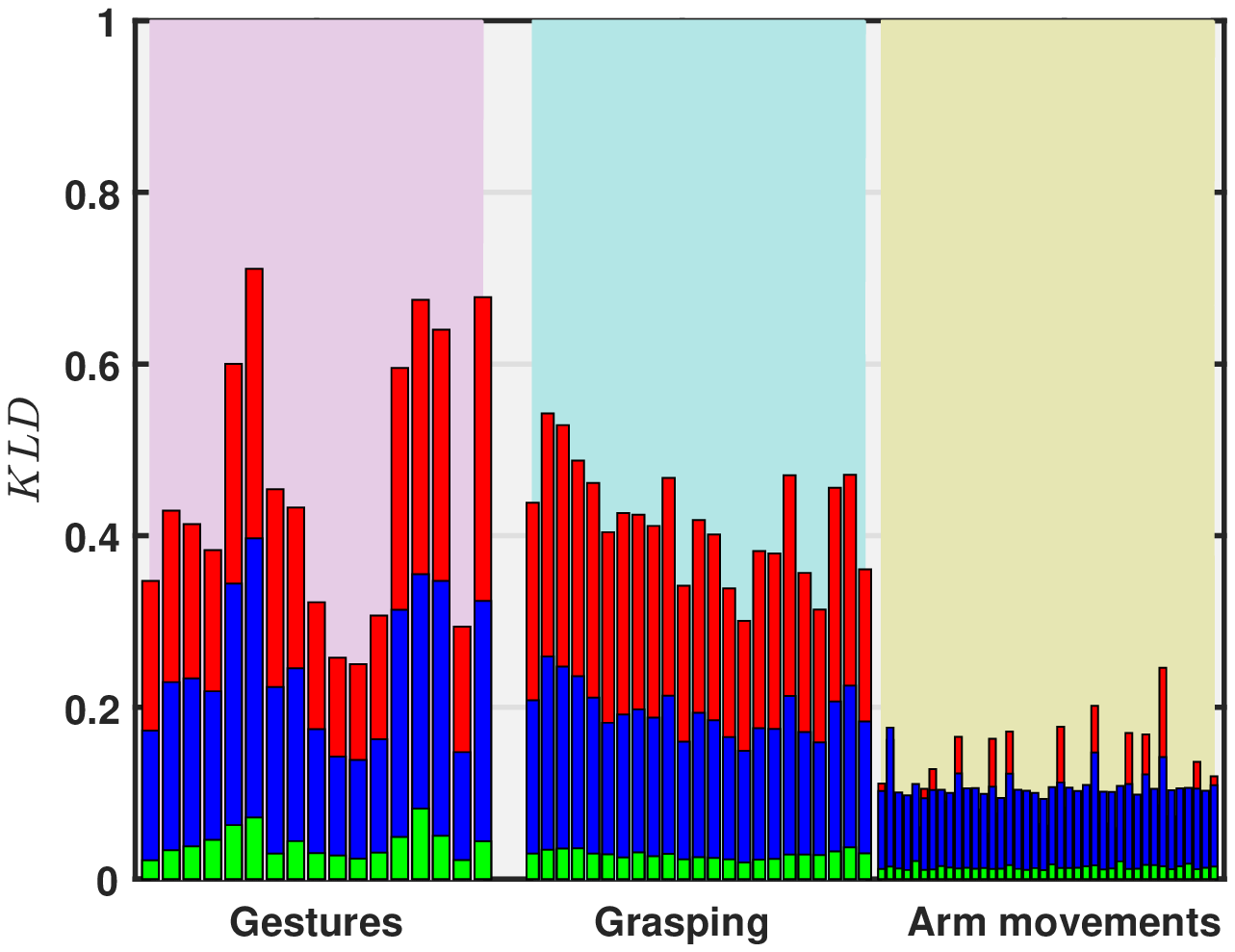}
          \caption{}
          \label{fig:NiceImage2}
      \end{subfigure}
      \caption{Average KLD for the 3 models (LGM-green, Laplacian-blue, Gaussian-red) for each of the trails (a) over the movements for different subjects (b) over the subjects for different movements }
      \label{Avg_KLD}
      \end{figure*}
      
\begin{figure*}[htb]
\captionsetup[subfigure]{justification=centering}
    \centering
      \begin{subfigure}{0.31\textwidth}
        \includegraphics[width=\textwidth]{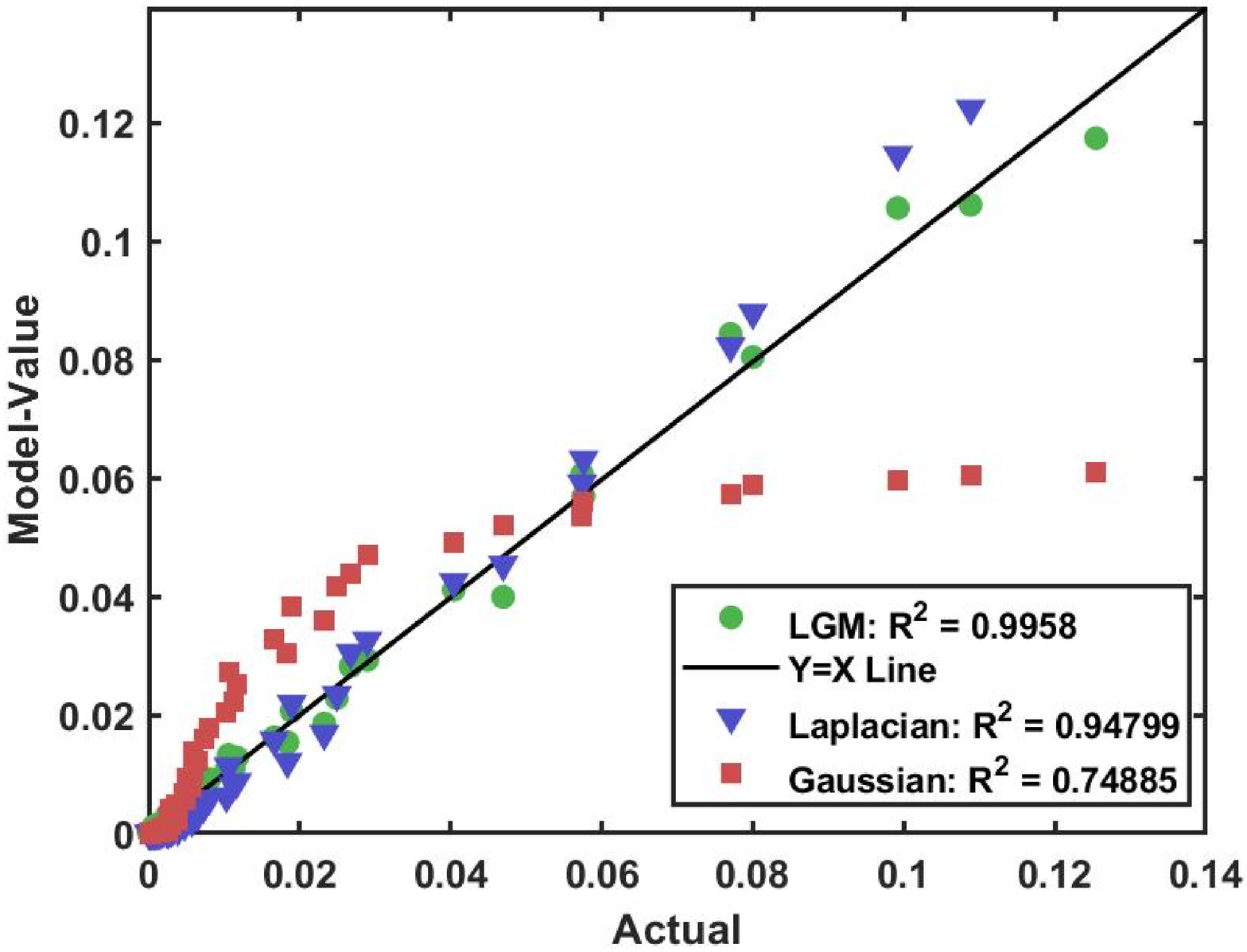}
          \caption{Gestures}
          \label{fig:NiceImage1}
      \end{subfigure}
      \begin{subfigure}{0.31\textwidth}
        \includegraphics[width=\textwidth]{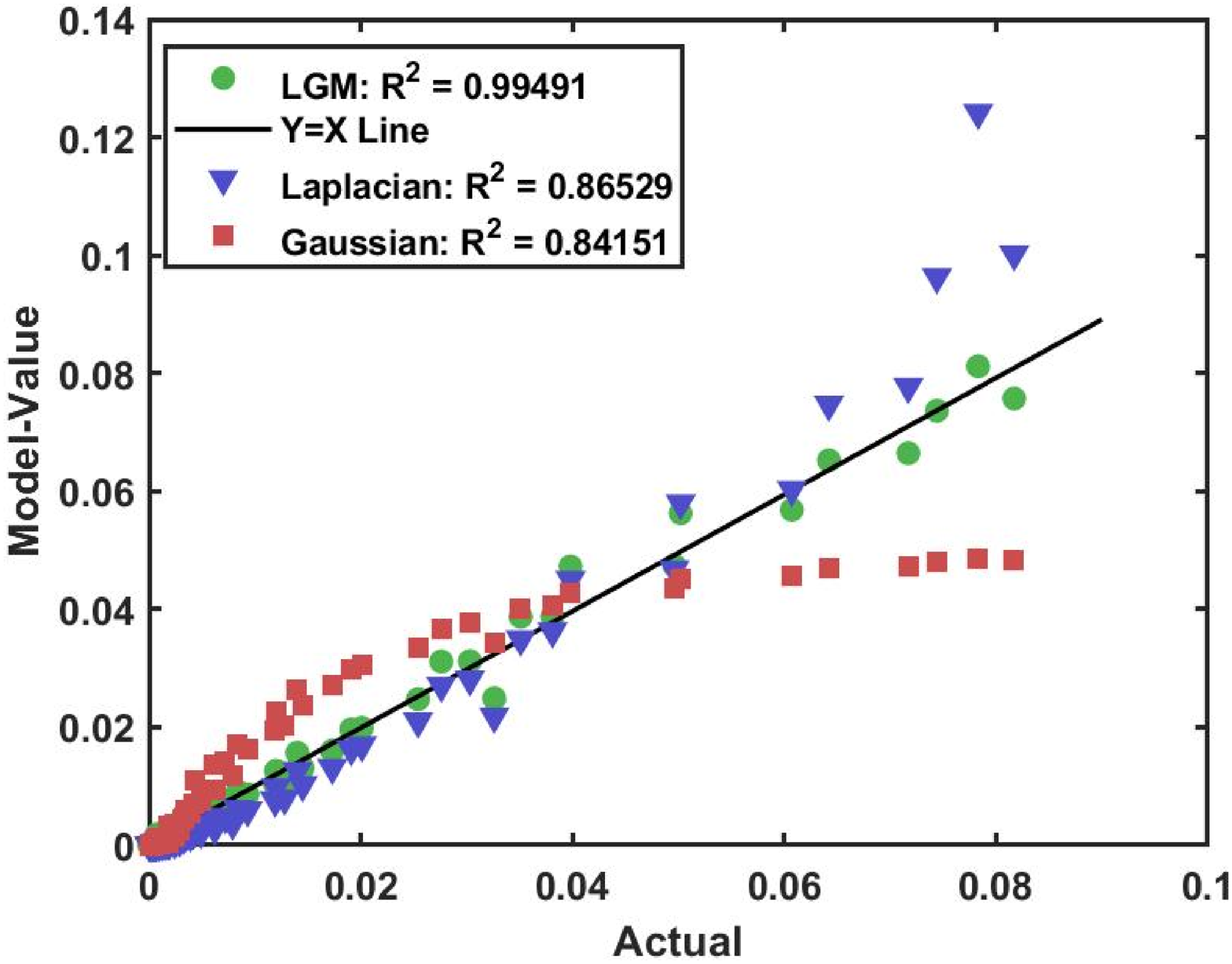}
          \caption{Grasping}
          \label{fig:NiceImage2}
      \end{subfigure}
      
      \captionsetup[subfigure]{justification=centering}
    \centering
      \begin{subfigure}{0.31\textwidth}
        \includegraphics[width=\textwidth]{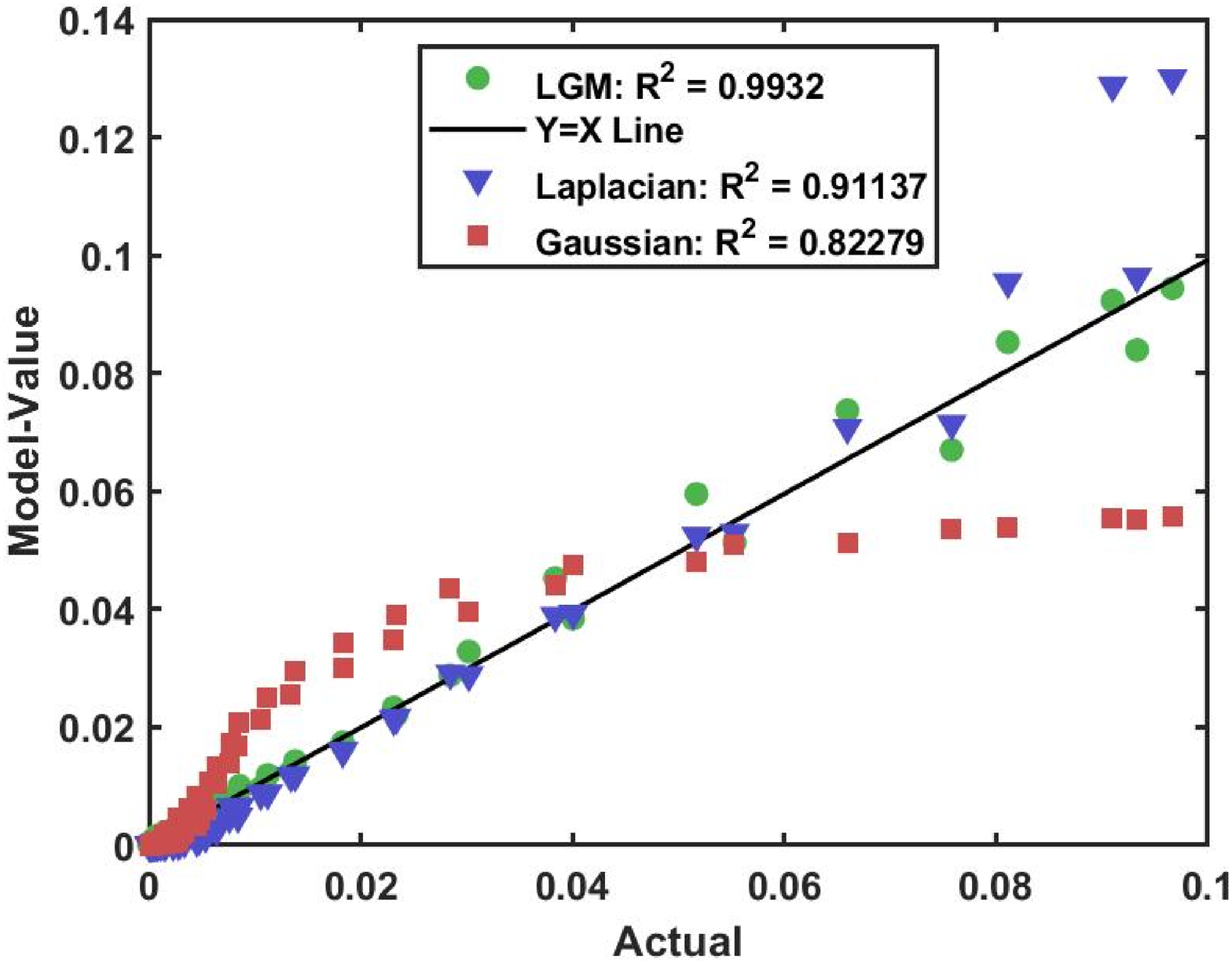}
          \caption{Normal activity}
          \label{fig:NiceImage1}
      \end{subfigure}
      \begin{subfigure}{0.31\textwidth}
        \includegraphics[width=\textwidth]{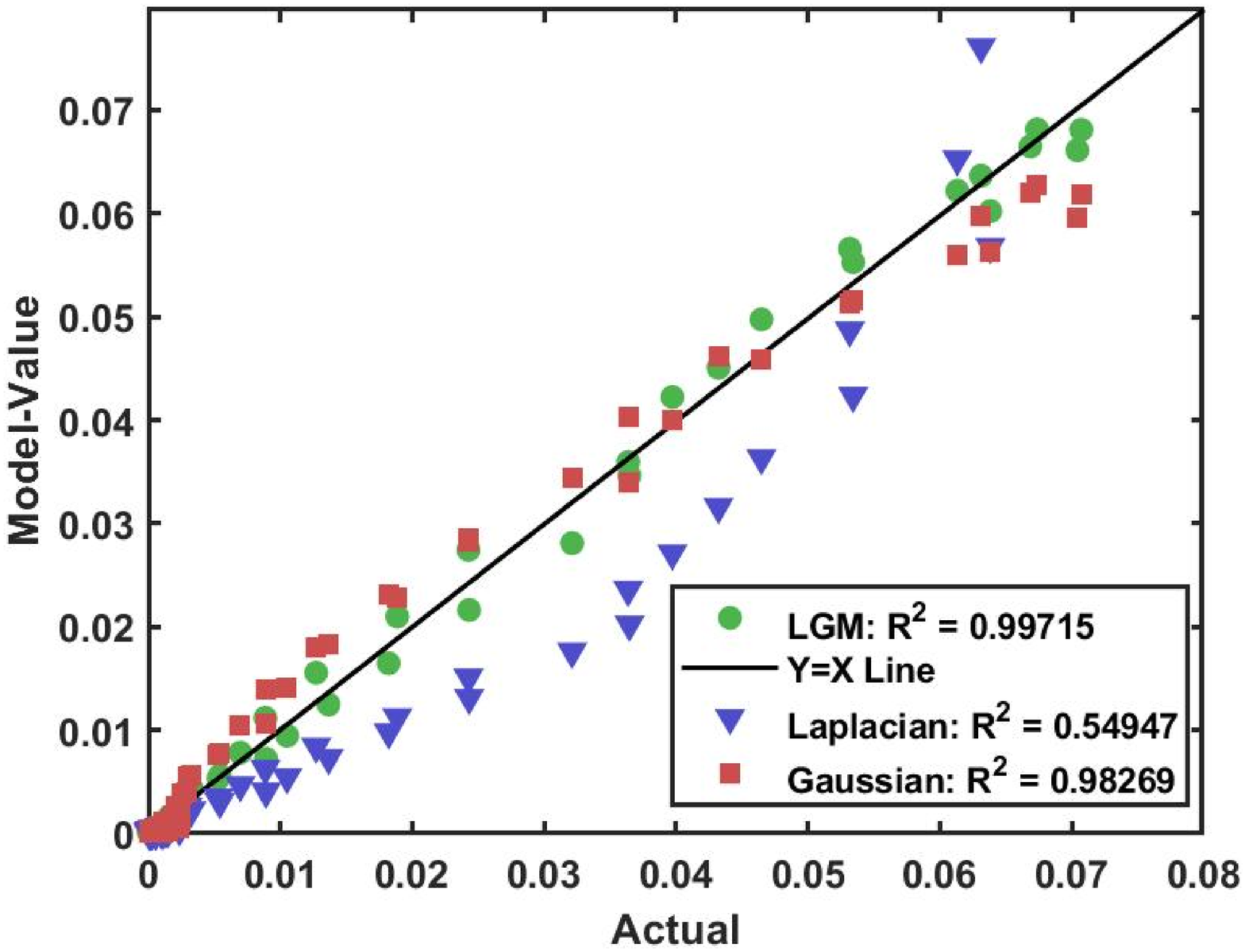}
          \caption{Intense activity}
          \label{fig:NiceImage2}
      \end{subfigure}
   
     \caption {Goodness of fit plots for the models LGM(green), Laplacian(blue) and Gaussian(red) for gestures, grasping, normal and Intense activities for the subjects-10, 3, 1 and 10 and with corresponding activities-7, 18, 3 and 1}
     \label{goodness}
      
\end{figure*}

\textbf{A goodness of fit plot with R-squared\cite{cohen1983applied}}: 
%\textcolor{blue}
{
The relationship between the sEMG data and the model-based values is analyzed using a goodness of fit plot. The nearer the data points are to the  line of equality, greater the model fit. On other hand, the coefficient of determination (R-squared) is a measure of how much the variance in the observed dependent variable is explained by the independent variable. The closer the value to $1$ greater the correlation between the two variables.}

%It can take any value between $0$ and $1$.

\textbf{Likelihood ratio test (LRT)}: 
{
The LRT is a statistical test used to compare two different models. In order to determine which model is statistically significant the likelihood values are evaluated for both the models. 
The LRT is defined as \cite{king1998unifying}
\begin{equation} \label{eq:LRT}
    T=2(log(L_p)-log(L_e))
\end{equation}
where $L_p$ and $L_e$ are likelihoods of the LGM model and {any existing model} respectively.}

% \begin{figure*}[h]
% \captionsetup[subfigure]{justification=centering}
%     \centering
%       \begin{subfigure}{0.31\textwidth}
%         \includegraphics[width=\textwidth]{hist_activity7_TBME.jpeg}
%           \caption{Gestures}
%           \label{fig:NiceImage1}
%       \end{subfigure}
%       \begin{subfigure}{0.31\textwidth}
%         \includegraphics[width=\textwidth]{hist_activity18_TBME.jpeg}
%           \caption{Grasping}
%           \label{fig:NiceImage2}
%       \end{subfigure}
     
%       \begin{subfigure}{0.31\textwidth}
%         \includegraphics[width=\textwidth]{hist_activity3_TBME.jpeg}
%           \caption{Arm activity}
%           \label{fig:NiceImage1}
%       \end{subfigure}
%       \begin{subfigure}{0.31\textwidth}
%         \includegraphics[width=\textwidth]{hist_activity34_TBME.jpeg}
%           \caption{Intense activity}
%           \label{fig:NiceImage2}
%       \end{subfigure}
      
%      \caption {Visual comparisons between mpdfs and estimated pdfs from models: LGM(green), Laplacian(blue) and Gaussian(red) for gestures, grasping, arm and intense activities for the subjects - 10, 3, 1 and  10 with corresponding activities - 7, 18, 3 and 1}
%       \label{fig:VisualComp}
% \end{figure*}
\section{Data Description}
Please note that all of the datasets analyzed in this study are available through public sources. 
Their short descriptions follow.  
%\textcolor{red}{ and these datasets are publicy avaliable}.
% All the datasets are publicly available.
\begin{itemize}
\item Ninapro Datasets:\\
In Ninapro DB2 (NPDB2) \cite{atzori2014electromyography} and DB4 (NPDB4) \cite{pizzolato2017comparison} datasets, there are $3$ exercises collected from groups of $40$ and $10$ subjects respectively. 
The exercises-$1$ and $2$ are related to activities such as hand gestures and grasping. The exercise-$3$ corresponds to finger movements at various forces levels including the abduction and adduction of the thumb. In this work, the EMG signals corresponding to the exercise- 2 from both the DB2 and DB4 are analyzed They consist of 23 grasping and 17 gesture actions respectively. The sEMG signals in this dataset have $12$ channels corresponding to a set of twelve electrodes placed at strategic muscle locations on an arm \cite{pizzolato2017comparison}. In this dataset, a typical sEMG signal within a activity, has a duration of $8$s with a $3$s rest time and $5$s activity. Each trial is repeated six times.

%Twelve electrodes were inserted at strategic muscle locations on the arm to collect EMG data for each of these muscles. . The EMG signal usually corresponds to 5s of recording with a 3s rest time for a given trial.}

\item Rami-khushaba DB6 (RKDB6) \cite{khushaba2014towards}:\\
%\textcolor{blue}
{
This dataset consists of sEMG signals collected from  11 intact subjects (9 males and 2 females) when they were performing $8$ different movements through $5$ limb positions. The limb positions were chosen in such a way that each subject can mimic daily activities. Each activity has six repetitions. A sEMG signal array consists of seven channels corresponding to seven Delsys DE $2.x$ EMG sensors placed across the circumference of the forearm}

\item Intense Action Dataset (IAD) \cite{ebied2020upper}:\\
%\textcolor{blue}
{
This dataset consists of sEMG signals acquired from $15$ healthy subjects when performing a single intense activity i.e., each subject is instructed to hold a 6kg dumbbell with the right hand for 120 seconds. These sEMG signals consist of $8$ channels corresponding to $8$ EMG electrodes}
{and each activity is carried out only once. The basic characteristics of these benchmark datasets are provided in the table \ref{Datasets summary}.}

\end{itemize}

\section{Results and Analysis} \label{sec:Results}
%\textcolor{blue}
{
For each of the mentioned datasets, the sEMG signals corresponding to each trial from each activity by each subject are analyzed using the three models. Specifically, the sEMG signal from the channel with the highest energy among multiple channels is examined using the models based on the following evaluation methods.   
\begin{itemize}
    \item  a qualitative analysis based on visual inspection
    \item  quantitative analyses: 
    \begin{enumerate}
    \item  the KL divergence analysis
    \item  the goodness of fit plots with R-squared and confidence interval for R-squared 
    \item  the likelihood ratio test
    \end{enumerate}
\end{itemize}
}
\subsection{{Visual Inspection}} 
%\textcolor{green}{The data sets were analysed with the models LGM, Laplacian and Gaussian for each of the subject and activity. The visual comparison plots corresponding to all subjects and activities are generated. }
%\textcolor{blue}
{
Fig. \ref{fig:VisualComp} illustrates the visual comparisons between the mpdf (yellow)  and the fitted pdfs from the LGM (green), the Laplacian (blue) and the Gaussian (red) models. These pdfs correspond to EMG signals of different activities as listed in the following: Fig. \ref{fig:VisualComp}(a): activity-$7$ i.e., pointing index finger by subject-$10$, Fig. \ref{fig:VisualComp}(b): activity-$18$ i.e., the quadpod grasp by subject-$3$, Fig. \ref{fig:VisualComp}(c): activity-$3$ i.e., a wrist supination by subject-$1$ and Fig. \ref{fig:VisualComp}(d): activity-$1$ i.e., lifting a dumbbell by subject-$10$.
Figs. \ref{fig:VisualComp}(a), (b) and (c) correspond to pdfs of the sEMG signal corresponding to gestures, grasping and normal arm activities. From these it is evident that the overlap between the mpdf and the LGM model is high compared to standalone Laplacian and Gaussian models. Whereas  Fig. \ref{fig:VisualComp}(d) represents the pdfs of the sEMG signal corresponding to the intense activity, it is noticed that the overlap between the LGM model and the  mpdf is similar to that of the standalone Gaussian model and the mpdf. In contrast, the overlap between the standalone Laplacian model and mpdf is lower. 
}

\subsection{Quantitative Analysis}
\subsubsection{KL-divergence}
%\textcolor{blue}

For each of the datasets under consideration, the KLD is evaluated between the LGM pdf and the mpdf. For comparison purposes, the KLD computation is also done for the Gaussian and the Laplacian pdfs against the mpdf. The corresponding results are illustrated in Figs. \ref{heatmap} to \ref{Avg_KLD}. Specifically, the heatmaps of KLD as a function of subjects and movements are shown in Fig. \ref{heatmap}. Each cell in a heatmap corresponds to the KLD for a given model for a particular subject while performing one of the activities. Further, the KLD represented here is an average over the given trials of an activity. Figs. \ref{heatmap} (a)-(c) correspond to the KLD for the Ninapro-DB4, Figs. \ref{heatmap} (d)-(f) depict the KLD for the Ninapro-DB2 and  Figs. \ref{heatmap} (g)-(i) represent the KLD for the Rami-khushaba-DB6. For each of the three datasets, it is noted that in these heatmaps, the LGM model has the lowest KLD. The lower and upper bounds of KLD for the heatmaps in Fig. 2 are shown in table \ref{tab:kldBound}. The key observation is the highest KLD value from the LGM model is the lowest KLD value for both the Laplacian and the Gaussian models.

 \begin{table}[h]
 \centering
 \caption{Lower and upper bounds of KLD for the proposed and the standalone models for different datasets}
\begin{tabular}{cccc}\\
\hline \hline
Datasets & LGM            & Laplacian      & Gaussian      \\
\hline NPDB4    & {[}0.01 0.1{]} & {[}0.1 0.6{]}  & {[}0.1 1.1{]} \\
\hline NPDB2    & {[}0.01 0.1{]} & {[}0.1 0.8{]}  & {[}0.1 2.2{]} \\
\hline RKDB6    & {[}0.01 0.1{]} & {[}0.1 0.35{]} & {[}0.1 0.9{]}\\
\hline \hline
\end{tabular}
\label{tab:kldBound}
\end{table}

{
The KLD values for different models in the case of the intense activity are shown in Fig. \ref{intense}. Notably, for the intense activity dataset as well, the KLD value is the lowest for the LGM model closely followed by the Gaussian model and then the Laplacian model. The minimum and maximum KLD values corresponding to the three models are: the LGM \{0.0033, 0.0301\}, the Gaussian \{0.0039, 0.0378\} and the Laplacian \{0.0920, 0.1513\}.
}
%{Clearly, the LGM model fits best to the sEMG signals under consideration.}} 
 
%We can further notice that the KLD values for the Gaussian model is close to that of the LGM model. However, the LGM model outperforms the Gaussian model.
%\textcolor{blue}
{
Fig. \ref{Avg_KLD}(a) shows the KLD averaged over the movements as a function of the subjects. Fig. \ref{Avg_KLD}(b) shows the vice versa case. The KLD of the LGM, Laplacian and Gaussian models are represented in green, blue and red respectively. From Fig. \ref{Avg_KLD}(a) and (b), it is observed that for the activities such as the gestures, grasping and the arm movements the average KLD value over the movements and subjects is the lowest for the LGM model, when compared to the other models.}

\subsubsection{Goodness of fit plots}
%\textcolor{blue}
{
Fig. \ref{goodness} illustrates the goodness of fit plots between the estimates from the three models versus the actual data. Specifically, the Figs. \ref{goodness} (a) to (d) correspond to the results on data from the gestures, grasping, normal arm and the intense activities respectively. The LGM model, Laplacian and Gaussian models are represented by the data points in green, blue and red respectively. From  Figs. \ref{goodness}  (a), (b) and (c), in the scatter plots, the model-values of the LGM model are found to be adjacent to the line of equality  which means that the predicted values from this model are close to the actual values of the sEMG signal.
Whereas in the case of intense activity shown in Fig. 5(d), the model values corresponding to both the LGM and the Gaussian models are similar and they are adjacent to the line of equality. From Fig. \ref{goodness}, for the gestures, grasping and normal arm activities, it can be concluded that the LGM model is better compared to other models. However, for the intense activity, both the LGM and the Gaussian fit the EMG data quite well. The average R-squared values are shown in the table \ref{tab:R2}. For the first three categories of activities, based on these metrics, the LGM model is found to be superior. Additionally, for the intense activities, the LGM and the Gaussian are again similar. The 95 percent confidence intervals(CI) \cite{finn1995correlations} \cite{cohen2014applied} for R-squared  corresponding to the plots in Fig. \ref{goodness} are given in table \ref{tab:CI}.
}
\begin{table}[h]
\centering
\caption{R-Squared values for four datasets from model evaluations}
\begin{tabular}{llll}\\
\hline \hline
Datasets & LGM     & Laplacain & Gaussian \\
\hline NPDB4    & 0.9958  & 0.94799   & 0.74885  \\
\hline NPDB2    & 0.99491 & 0.86529   & 0.84151  \\
\hline RKDB6    & 0.9932  & 0.91137   & 0.82279  \\
\hline IAD      & 0.99715 & 0.54947   & 0.98269 \\
\hline \hline
\end{tabular}
\label{tab:R2}
\end{table}

 \begin{table}[h]
 \centering
 \caption{Confidence interval of R-squared for LGM, Laplacian and Gaussian models}
\begin{tabular}{llll}\\
\hline \hline
\hline Datasets & LGM                 & Laplacain           & Gaussian            \\
\hline NPDB4    & {[}0.9946 0.9970{]} & {[}0.9439 0.9520{]} & {[}0.7413 0.7564{]} \\
\hline NPDB2    & {[}0.9937 0.9961{]} & {[}0.8598 0.8708{]} & {[}0.8357 0.8473{]} \\
\hline RKDB6    & {[}0.9918 0.9946{]} & {[}0.9067 0.9160{]} & {[}0.8167 0.8289{]} \\
\hline IAD      & {[}0.9960 0.9983{]} & {[}0.5402 0.5587{]} & {[}0.9800 0.9854{]}\\
\hline \hline
\end{tabular} \label{tab:CI}
\end{table}

\subsubsection{Likelihood ratio test}
%\textcolor{blue}
{
The LRT given in (\ref{eq:LRT}) is carried out between the LGM model and the Laplacian model as shown below. 
\begin{eqnarray}
H_0: & \text{The Laplacian model fits the data} \nonumber \\
H_1: & \text{The LGM model fits the data}   \nonumber 
\end{eqnarray}
This test is carried out for $99\%$ confidence interval. It is noted that p-value is less than  $0.01$, which means $H_0$ is rejected and $H_1$ is accepted. The test is repeated by replacing the Laplacian with the Gaussian model for $H_0$. In this test also the p-value is found to be less than $0.01$ and thus $H_1$ is accepted. 
}

%\textcolor{orange}{Let $H_0$ be the null hypothesis, the data come from Laplacian model and $H_1$ denotes alternative hypothesis, the data come from LGM model.}
% \begin{table}[h]
% \begin{tabular}{|c|c|c|}
% \hline
% \textbf{Activities} & \textbf{Level of significance} & \textbf{Decision} \\ \hline
% Gestures             & 99\%                           & $H_o$ is rejected  \\ \hline
% Grasping            & 99\%                           & $H_o$ is rejected  \\ \hline
% Arm activities   & 99\%                           & $H_o$ is rejected  \\ \hline
% Intense activities  & 99\%                           & $H_o$ is rejected  \\ \hline
% \end{tabular}
% \end{table}

% \begin{table}[h] 
% \centering
% \caption{R-Squared values for different datasets from the model evaluations}  
% \begin{tabular}{|l|c|c|c|}
% \hline
%  \multicolumn{1}{|l|}{Datasets}& \multicolumn{1}{l|}{LGM} & \multicolumn{1}{l|}{Laplacian} & \multicolumn{1}{l|}{Gaussian} \\ \hline
% \multicolumn{1}{|c|}{NPDB4} & 0.9958                   & 0.94799                        & 0.74885                       \\ \hline
% NPDB2                       & 0.99491                  & 0.86529                        & 0.84151                       \\ \hline
% RKDB6                 & 0.9932                   & 0.91137                        & 0.82279                       \\ \hline
% IAD set                  & 0.99715                  & 0.54947                       & 0.98269                       \\ \hline
% \end{tabular}  \label{tab:R2}
% \end{table}

\begin{figure}[t]
        \centering
      \includegraphics[width=1.00\columnwidth]{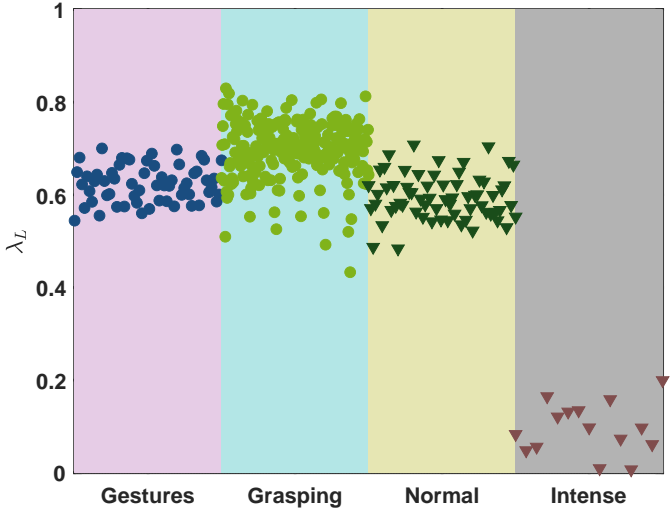}
      \caption{Laplacian coefficient for each trail versus subjects for gestures, grasping, normal arm movements and intense activity}
     \label{mixing}
\end{figure}

\begin{figure}[t!]
\captionsetup[subfigure]{justification=centering}
    \centering
      \begin{subfigure}{0.31\textwidth}
        \includegraphics[width=\textwidth]{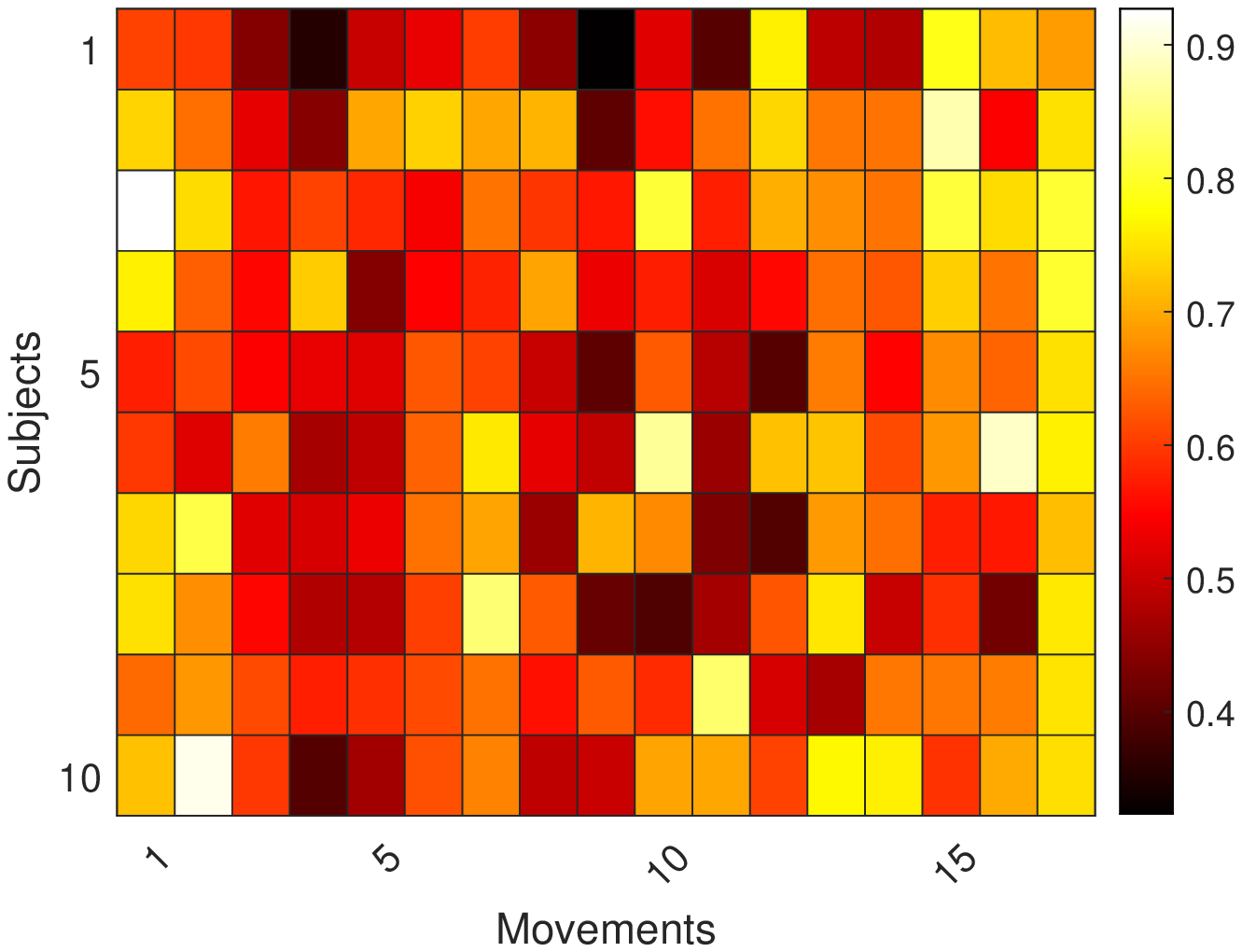}
          \caption{}
          \label{fig:NiceImage1}
      \end{subfigure}
      \begin{subfigure}{0.31\textwidth}
        \includegraphics[width=\textwidth]{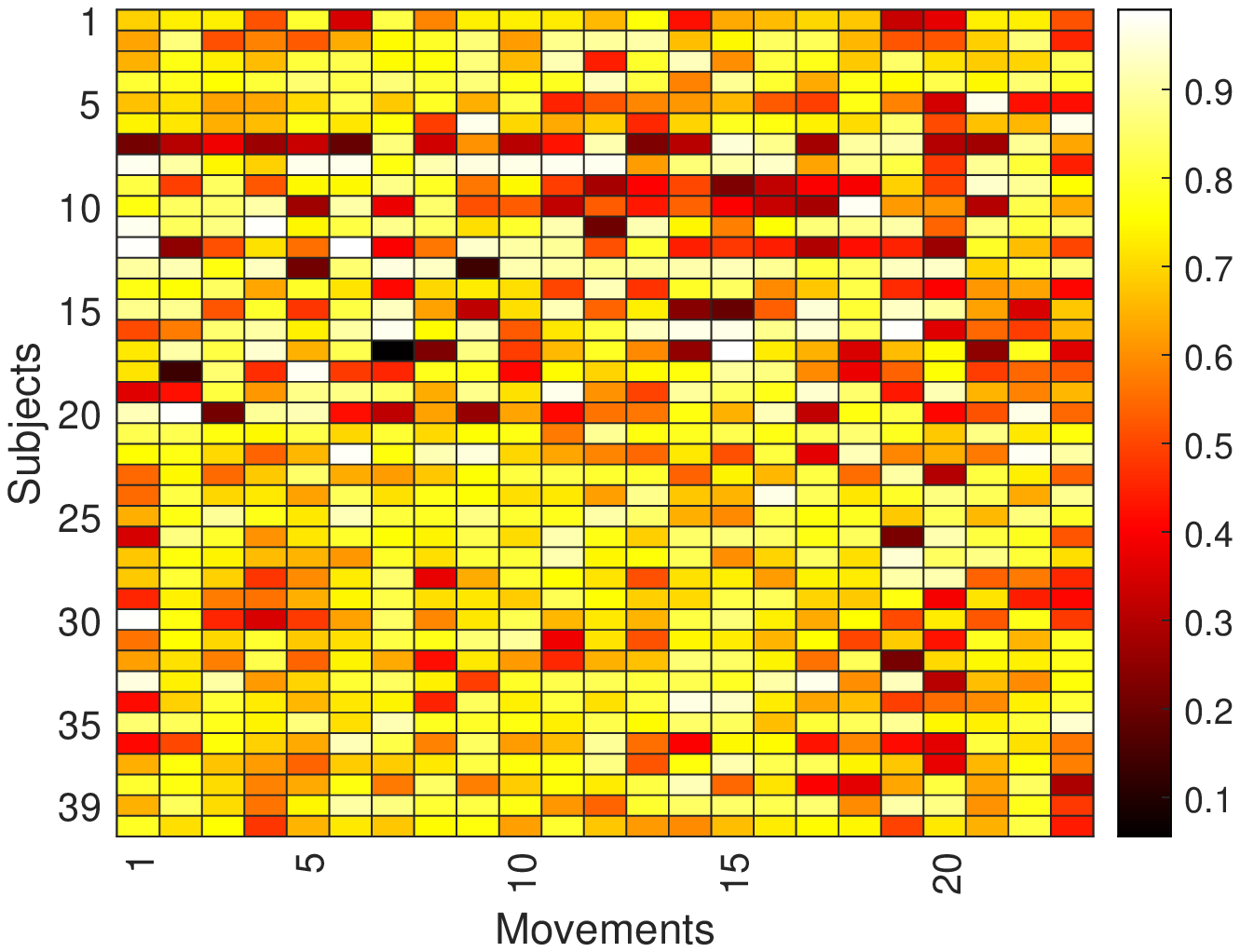}
          \caption{}
          \label{fig:NiceImage2}
      \end{subfigure}
      \begin{subfigure}{0.31\textwidth}
        \includegraphics[width=\textwidth]{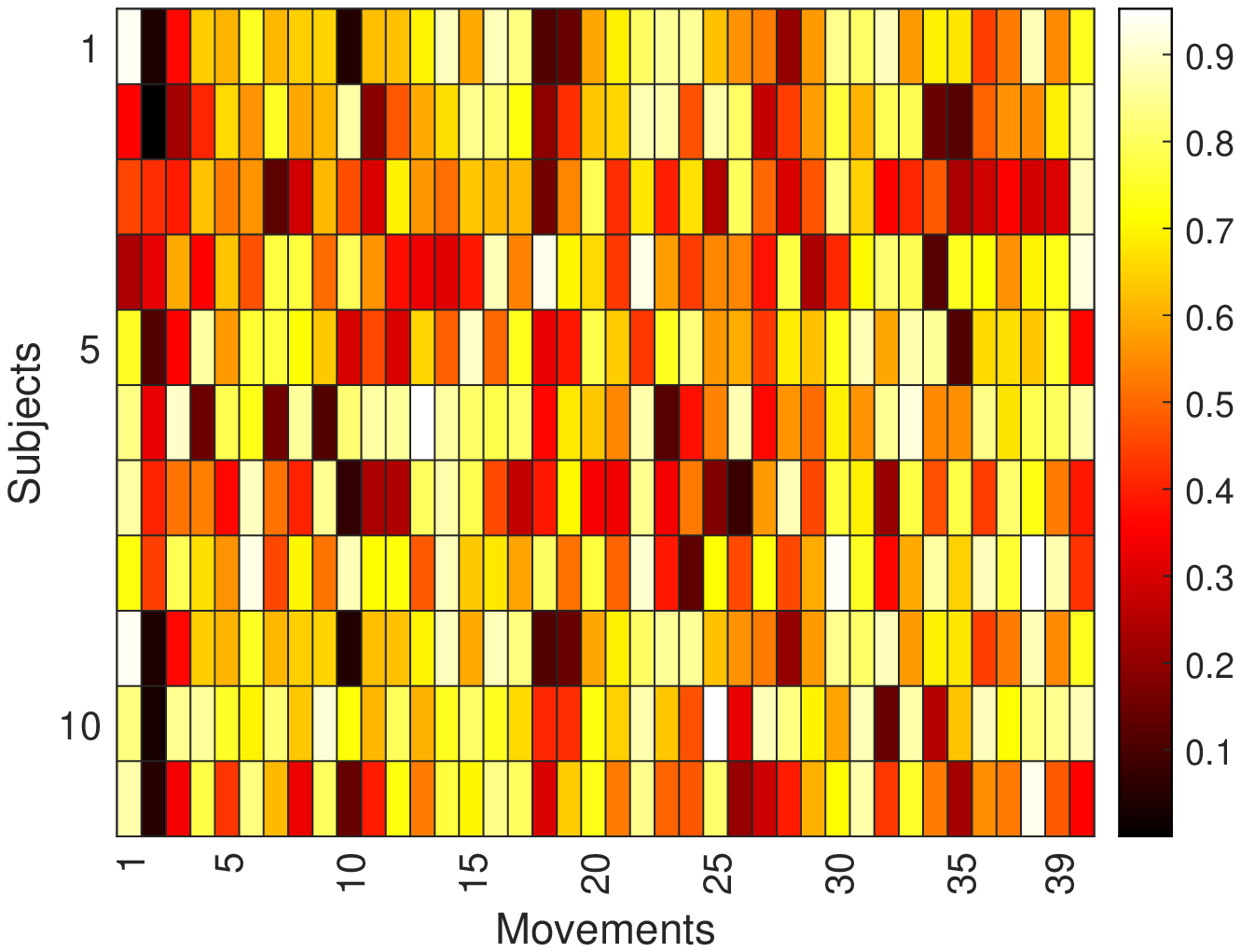}
          \caption{}
          \label{fig:NiceImage2}
      \end{subfigure}
      \caption{Laplacian weights of LGM model averaged over trials as a function of subjects and movements for (a) gestures, (b) grasping, and (C) arm activities }
      \label{Laplacian Weights}
      \end{figure}

\subsection{Mixing Weights}
%\textcolor{blue}
{
The mixing coefficients corresponding to the Laplacian component of the LGM model corresponding to different activities from various datasets are shown in Fig. \ref{mixing}. 
For the low and medium MCF levels such as the gestures, grasping and the normal activities, it is noticed that the Laplacian component has a stronger weighting in comparison with the Gaussian component.  Whereas in case of intense activity i.e., at high MCF level the mixing weight corresponding to the Gaussian component is higher and the Laplacian component is lower.
Thus from Fig. \ref{mixing}, it can be recognized that as the intensity of an activity, i.e., the amount of energy required for {performing} a certain action increases, the weight of the Laplacian component reduces. Fig. \ref{Laplacian Weights} shows Laplacian weights of the LGM model corresponding to the first three limb activities under consideration. The horizontal and vertical axes in the heatmaps correspond to the number of movements and subjects respectively. Each cell in a heatmap denotes the Laplacian weight in the LGM model for a particular subject and activity. It is noticed that for most of the cases the Laplacian weight $\lambda_{1}$ is dominating the Gaussian weight $\lambda_{2}$. In some circumstances, the Laplacian weights are lower than Gaussian weights. For example, in Fig. \ref{Laplacian Weights} (a) for the subject-$1$, activities-$4$ and $9$, in Fig. \ref{Laplacian Weights}(b) for the subject-$18$, activity-$2$ and in Fig. \ref{Laplacian Weights} (c), the subject-$1$, activities-$2$ and $10$, the Gaussian weights are stronger.}

\section{Discussion}
%\subsection{Discussion}
%\textcolor{blue}
{
From the results presented in section \ref{sec:Results}, for the EMG signals corresponding to the low and medium levels of muscle recruitment i.e., for activities such as the gestures, grasping and the normal arm movements, the LGM is found to be a more suitable model compared to the standalone models. This is verified in terms of 1) the visual inspection between a model pdf and the mpdf, 2) the lowest KLD, 3) the goodness of fit plots - the model values matching the true values, 4) the higher R-squared values and 5) the Likelihood ratio test accepting the alternate hypothesis.
However, in the case of intense activities, both the LGM and the Gaussian model seem to perform quite similarly according to the four evaluation methods described above. Hence for high levels of muscle recruitment, the proposed LGM model behaves similar to a standalone Gaussian model. This result is further qualified by the following observation, in the LGM model, the mixing coefficient of the Laplacian component becomes very small in comparison to that of the Gaussian component.}

%\color{blue}
{In the analysis on mixing weights, it is observed that for the first three types of actions, both the Laplacian and Gaussian components have significant contributions to the model. However, for the intense activity, the Gaussian component is much stronger. Hence, from these findings it can be postulated that the weights of the LGM model can be related to MCF level and motor units that are activated during an activity. 
For example, for the first three activities, the Laplacian weight $\lambda_1$ is higher relating to lower MCF level and the lower number of activated motor units. However, in the case of intense activity the Gaussian weight $\lambda_2$ is higher connecting to a higher MCF level and a larger number of activated motor units. These findings are in agreement with the literature on pdfs reported in section \ref{sec:LitRev} where it is noted that for the lower and medium MCF levels, the pdfs have a sharper peak at center, hinting a Laplacian structure, and at higher MCF levels, they have a clear Gaussian structure.
}
\section{Conclusion}

In this paper, a Laplacian Gaussian mixture model is proposed for sEMG signals from upper limbs. The proposed model is tested on several benchmark sEMG datasets and compared with the existing standalone models. 
The suitability of the model is validated using (1) qualitative analyses such as visual comparison with the empirical pdf (mpdf) where it is observed that the LGM model has the best agreement, (2) the KL divergence between the model pdf and the mpdf, again the KLD is lowest for the LGM model, (3) a goodness of fit plot, comparison of coefficient of determination (CFD) - $R^2$ and confidence intervals for $R^2$, here it is noted that $R^2$ in case of the LGM model is closest to unity and (4) the Likelihood ratio test (LRT) that also supported the LGM model. Finally, it is noted, for the low and medium muscle contraction force levels, the Laplacian weight has stronger weighting than the Gaussian. Whereas for the higher muscle contraction force levels the Laplacian weights are lower. {In the future work, we will extend the proposed model to understand the correlations between the sEMG signals from various muscle locations.}
  
\section*{Acknowledgment}
This research is funded by SERB, Govt. of India under Project Grant No. CRG/2019/003801.

\bibliographystyle{IEEEtran}
\bibliography{References.bib}

% \newpage

% \section{Biography Section}
% If you have an EPS/PDF photo (graphicx package needed), extra braces are
%  needed around the contents of the optional argument to biography to prevent
%  the LaTeX parser from getting confused when it sees the complicated
%  $\backslash${\tt{includegraphics}} command within an optional argument. (You can create
%  your own custom macro containing the $\backslash${\tt{includegraphics}} command to make things
%  simpler here.)
 
% \vspace{11pt}

% \bf{If you include a photo:}\vspace{-33pt}
% \begin{IEEEbiography}[{\includegraphics[width=1in,height=1.25in,clip,keepaspectratio]{fig1}}]{Michael Shell}
% Use $\backslash${\tt{begin\{IEEEbiography\}}} and then for the 1st argument use $\backslash${\tt{includegraphics}} to declare and link the author photo.
% Use the author name as the 3rd argument followed by the biography text.
% \end{IEEEbiography}

% \vspace{11pt}

% \bf{If you will not include a photo:}\vspace{-33pt}
% \begin{IEEEbiographynophoto}{John Doe}
% Use $\backslash${\tt{begin\{IEEEbiographynophoto\}}} and the author name as the argument followed by the biography text.
% \end{IEEEbiographynophoto}

% \vfill

\end{document}